\documentclass[numberedappendix]{emulateapj}

\shorttitle{SLACS I: A Large Sample of Lens Galaxies }
\shortauthors{Bolton et al.\ }

\newcommand{\oii}{\mbox{[O~{\sc ii}]} 3727}

\newcommand{\oiiib}{\mbox{[O~{\sc iii}]} 5007}

\newcommand{\hb}{H$\beta$}

\usepackage{natbib}
\usepackage{graphicx}

\begin{document}

\bibliographystyle{apj}

\title{The Sloan Lens ACS Survey. I\@. A Large 
Spectroscopically Selected Sample of Massive
Early-Type Lens Galaxies\altaffilmark{1}}

\author{Adam~S.~Bolton\altaffilmark{2,3},
 Scott Burles\altaffilmark{2},
 L\'{e}on~V.~E.~Koopmans\altaffilmark{4},
 Tommaso~Treu\altaffilmark{5,6},
 and Leonidas~A.~Moustakas\altaffilmark{7}}

\altaffiltext{1}{Based on data from the Sloan Digital Sky Survey,
 the 6.5-m Walter Baade (Magellan-I) Telescope, the 8-m Gemini-North
 Telescope, and the {\sl Hubble Space Telescope}.  See
 full acknowledgement at end.}
 \altaffiltext{2}{Department of Physics and Kavli Institute for
 Astrophysics and Space Research, Massachusetts Institute of
 Technology, 77 Massachusetts Ave., Cambridge, MA 02139, USA ({\tt
 bolton@alum.mit.edu, burles@mit.edu})}
 \altaffiltext{3}{Harvard-Smithsonian Center for Astrophysics,
 60 Garden St., Cambridge, MA 02138, USA ({\tt abolton@cfa.harvard.edu})}
 \altaffiltext{4}{Kapteyn
 Astronomical Institute, University of Groningen, P.O. Box 800, 9700AV
 Groningen, The Netherlands ({\tt koopmans@astro.rug.nl})}
 \altaffiltext{5}{Department of Physics, University of California,
 Santa Barbara, CA 93101, USA ({\tt tt@physics.ucsb.edu})}
 \altaffiltext{6}{Department of Physics and Astronomy, UCLA, Box
 951547, Knudsen Hall, Los Angeles, CA 90095, USA}
 \altaffiltext{7}{Jet Propulsion Laboratory/Caltech, MS 169-327, 4800
 Oak Grove Dr., Pasadena, CA 91109, USA ({\tt leonidas@jpl.nasa.gov})}

\slugcomment{ApJ, in press}

\begin{abstract}
The Sloan Lens ACS (SLACS) Survey is an efficient {\em Hubble Space
Telescope} ({\sl HST}) Snapshot imaging survey for new galaxy-scale
strong gravitational lenses.  The targeted lens candidates are
selected spectroscopically from within the Sloan Digital Sky Survey
(SDSS) database of galaxy spectra for having multiple nebular emission
lines at a redshift significantly higher than that of the SDSS target
galaxy.  The SLACS survey is optimized to detect bright early-type
lens galaxies with faint lensed sources, in order to increase the
sample of known gravitational lenses suitable for detailed lensing,
photometric, and dynamical modeling.  In this paper, the first in a
series on the current results of our {\sl HST} Cycle--13 imaging
survey, we present a catalog of 19 newly discovered gravitational
lenses, along with 9 other observed candidate systems that are either
possible lenses, non-lenses, or non-detections.  The survey efficiency
is thus $\ge 68$\%.  We also present Gemini 8m and Magellan 6.5m
integral-field spectroscopic data for 9 of the SLACS targets, which
further support the lensing interpretation.  A new method for the
effective subtraction of foreground galaxy images to reveal faint
background features is presented.  We show that the SLACS lens
galaxies have colors and ellipticities typical of the spectroscopic
parent sample from which they are drawn (SDSS luminous red galaxies
and quiescent main-sample galaxies), but are somewhat brighter and
more centrally concentrated. Several explanations for the latter bias
are suggested.  The SLACS survey provides the first statistically
significant and homogeneously selected sample of bright early-type
lens galaxies, furnishing a powerful probe of the structure of
early-type galaxies within the half-light radius.  The high
confirmation rate of lenses in the SLACS survey suggests consideration
of spectroscopic lens discovery as an explicit science goal of future
spectroscopic galaxy surveys.
\end{abstract}


\section{INTRODUCTION}

In the currently favored cosmological scenario, the matter content of
the universe is dominated by a cold and dark component of unknown
particle species whose only significant interaction with the smaller
baryonic matter fraction occurs through the gravitational force.  This
``cold dark matter'' (CDM) picture is most strongly required by
observations on the largest scales \citep{spergel_wmap, tegmark_sdss,
percival_2df}.  The CDM scenario holds that galaxies form within the
potential wells of extended dark-matter halos which began their
collapse while baryonic matter was still ionized \citep{white_rees_78,
blumenthal_84}.  This view is supported by direct evidence for dark
matter in disk galaxies from the observation of rotational velocities
that remain approximately constant out to radii at which the stellar
galactic component makes a diminishing contribution
\citep[e.g.][]{Rubin_1980, van_albada_sancisi}.

Unlike disk galaxies, early-type galaxies (E and S0) are generally
pressure supported and lack bright kinematic tracers at large radii
such as H$_{\scriptsize{\rm I}}$. As such their kinematics are more
difficult to measure and interpret in terms of mass density profiles.
The density structure of early-type galaxies is nevertheless of great
interest for numerous reasons.  First, their structure is the ``fossil
record'' of their formation and evolution processes
\citep[e.g.][]{wechsler_02, zhao_03, loeb_peebles, gao_2004}.
Hierarchical CDM galaxy-formation theories hold that early-type
galaxies are built through the merging of late-types \citep*{kwg_1993,
bcf_1996}, which should have predictable consequences for the
structure of the merger products.  The most stringent test of these
theories will require precise observational measurements of early-type
mass profiles.  For example, detailed measurement of the structure of
high-surface-brightness early-type galaxies will enable quantitative
tests of the CDM theory on scales where baryonic and radiative
processes have significant effects upon the structure of the host
dark-matter halo \citep[e.g.\ through ``adiabatic contraction'',
e.g.][]{flores_adiabat, blumenthal_adiabat, mo_mao_white,
gnedin_adiabat}, altering it significantly relative to the form
expected to result from pure collisionless dark-matter collapse
\citep[e.g.][]{nfw, moore_98}.  Second, early-type galaxies exhibit
great regularity in their photometric, spectroscopic, and kinematic
properties, as described, e.g., by the well-known Fundamental Plane
(FP) relation between velocity dispersion, effective radius, and
surface brightness \citep{dd_fp, dr_fp}.  The ``tilt'' of the FP
relative to the simple expectation of the virial theorem can be
understood in terms of a dependence of galaxy structure and
mass-to-light ratio on total galaxy mass \citep{bbf_92, clr_96,
bcd_02, tbb_04}.  However, further constraints on the mass structure
of early-type galaxies are needed in order to distinguish between the
various effects of differing stellar populations
(the explanation given by, e.g., \citealt{gerhard_2001}),
differing density
profiles (``structural nonhomology'' or ``weak homology''), and
differing dark-matter fractions
(the explanation of, e.g., \citealt{padmanabhan_04})
in giving rise to the FP.

Despite great progress, observational results remain uncertain because
of the small number of individual galaxies suitable for study.
Stellar-dynamical measurements of local early-type galaxies
\citep[e.g.][]{bertin_94, franx_94, gerhard_2001, capp_sauron}, the
statistics of early-type gravitational-lens galaxies
\citep[e.g.][]{kochanek_lensstat_96, chae_lensstat_03, rusin_lensstat,
rusin_kochanek_05}, and combined lensing and dynamical measurements of
the few systems amenable to such study \citep[][hereafter
KT]{kt02,kt03,tk02,tk03,tk04} generally argue for the presence of a
significant amount of dark matter even on the scale of the half-light
radius, leading to an approximately linear increase of enclosed mass
with radius and thus to a flat equivalent rotation curve as in disk
galaxies.  However, in some cases the presence of dark matter is not
required, and the observed kinematics can be reproduced with a
constant mass-to-light ratio \citep[e.g.][]{bertin_94, romanowsky_pn},
corresponding to a declining rotation curve. Furthermore,
\citet{kochanek_cdm_delay} has pointed out an apparent conflict
between isothermal mass profiles (i.e.\ flat rotation curves) and some
gravitational-lens time delays under the assumption of $H_0 \approx$
70~km~s$^{-1}$~Mpc$^{-1}$. Due to this persistent uncertainty about
the structure and diversity of early-type galaxies \citep[see
also][]{kochanek_where}, it is important to collect precision data for
a larger number of objects, possibly spanning a large range of
redshifts, environments, and mass.

Strong gravitational lensing provides the most direct probe of mass in
early-type galaxies: a measurement of the mass enclosed within the
Einstein radius.  When combined with a spatially resolved measurement
of the line-of-sight velocity-dispersion profile and the surface
brightness of the lens galaxy, this mass measurement can be used to
constrain the luminous and dark-matter mass profiles simultaneously
through the Jeans equation (e.g.\ KT).  Unfortunately, only a handful
of known strong lenses are amenable to this type of analysis, often
because the back-ground source is a bright QSO whose images outshine
the lens galaxy.

We have therefore initiated the Sloan Lens ACS (SLACS) Survey
\citep[see][]{bolton_1402} to discover a much larger sample of new
early-type strong gravitational lenses suitable for detailed
photometric and dynamical study and thereby realize the full potential
of gravitational lensing as a probe of early-type galaxy structure.
SLACS uses the method described by \citet[][hereafter
B04]{bolton_speclens} to select candidate galaxy-scale gravitational
lens systems from within the Sloan Digital Sky Survey (SDSS)
spectroscopic database on the basis of multiple higher-redshift
emission lines in the spectrum of lower-redshift target galaxies.
For discussion of other spectroscopic lens surveys and
lens discoveries based on the identification of
anomalous emission lines, see \citealt{huch85, warren_0047_96,
warren_0047_98, warren_0047_99, wil00, hall00, h00, john03, willis_05}.
The SDSS data also provide stellar velocity dispersion measurements for
all SLACS lens candidates.  The most promising candidates (see
\S~\ref{sec:selection}) are observed through F435W and F814W using the
Advanced Camera for Surveys (ACS) aboard the {\sl Hubble Space
Telescope} ({\sl HST})\@.  ACS images enable us to confirm bona fide
lenses, measure detailed photometric and morphological parameters of
the lensing galaxies, and obtain accurate astrometry and
surface-brightness measurements of strongly lensed features with which
to constrain gravitational-lens models.  Finally, confirmed lenses
from the program are being targeted for deep ground-based spectroscopy
in order to obtain spatially resolved velocity-dispersion profiles of
the lensing galaxies, which will provide more precise dynamical
constraints on their mass structure.  

This paper (Paper I) is the first in a series presenting the results
of the current Cycle--13 {\sl HST} imaging component of the SLACS
Survey.  This paper presents the catalog of new lenses confirmed by
the SLACS {\sl HST}-ACS Snapshot Survey and their properties as
measured by SDSS, describes our methods of image processing and
lens-galaxy image subtraction, and provides an analysis of the
possible selection biases that bear upon whether or not our lenses are
a representative sample of similar SDSS early-type galaxies.  We also
present ground-based integral-field spectroscopy of several SLACS
systems which supports a strong-lensing interpretation of the observed
features.  Paper II presents photometric and morphological
measurements of the SLACS lens sample from ({\sl HST}) imaging and
places the sample within the context of the FP, Paper III presents
gravitational-lens and dynamical modeling results, and Paper IV
focuses on the properties of the lensed source galaxies.

The outline of this paper is as follows: In \S\,\ref{sec:survey}, the
survey selection procedure and the available data of each system is
described. In \S\,\ref{lenscat}, nineteen newly discovered lens
systems and nine unconfirmed systems or non-detections are
presented. In \S\,\ref{sec:stats}, the statistics and possible
selection effects of the sample are discussed. \S\,\ref{sec:future}
discusses some of the implications of the SLACS survey for future
surveys and in \S\,\ref{sec:conclusions} conclusions are drawn and
future work is discussed. Throughout this paper, we assume a
cosmological model with $\Omega_M = 0.3$, $\Omega_\Lambda = 0.7$, and
$H_0 = 70 h_{70}$\,km\,s$^{-1}$\,Mpc$^{-1}$ (with $h_{70} = 1$).

\section{THE SURVEY}\label{sec:survey}

\begin{deluxetable*}{lccccccccc}
\tabletypesize{\scriptsize}
\tablewidth{\hsize}
\tablecaption{Properties of Observed Systems}
\tablehead{
\colhead{System Name} & 
\colhead{Plate-MJD-Fiber} & 
\colhead{$g,r,i$} & 
\colhead{$R_{\mathrm{eff}}$ ($\arcsec$)} & 
\colhead{$z_{\mathrm{FG}}$} & 
\colhead{$z_{\mathrm{BG}}$} & 
\colhead{$\sigma_a$ (km\,s$^{-1}$)} & 
\colhead{Sample} &
\colhead{Lens}} 
\startdata

SDSS J003753.21$-$094220.1 & 
0655-52162-392 & 
18.00,16.81,16.39 & 
$2.16 \pm 0.06$ & 
0.1954 & 
0.6322 & 
$265 \pm 10$ & 
LRG &
Yes\\ 

SDSS J021652.54$-$081345.3 & 
0668-52162-428 & 
19.07,17.46,16.90 & 
$3.05 \pm 0.13$ & 
0.3317 & 
0.5235 & 
$332 \pm 23$ & 
LRG &
Yes\\ 

SDSS J073728.45$+$321618.5 & 
0541-51959-145 & 
19.38,17.84,17.15 & 
$2.16 \pm 0.13$ & 
0.3223 & 
0.5812 & 
$310 \pm 15$ & 
LRG &
Yes\\ 

SDSS J081931.92$+$453444.8 & 
0441-51868-108 & 
18.63,17.51,17.07 & 
$2.32 \pm 0.13$ & 
0.1943 & 
0.4462 & 
$231 \pm 16$ & 
MAIN &
?\\ 

SDSS J091205.30$+$002901.1 & 
0472-51955-429 & 
17.31,16.22,15.78 & 
$3.36 \pm 0.05$ & 
0.1642 & 
0.3239 & 
$313 \pm 12$ & 
LRG &
Yes\\ 

SDSS J095320.42$+$520543.7 & 
0902-52409-577 & 
18.57,17.61,17.22 & 
$1.77 \pm 0.09$ & 
0.1310 & 
0.4670 & 
$207 \pm 14$ & 
MAIN &
?\\ 

SDSS J095629.77$+$510006.6 & 
0902-52409-068 & 
18.41,17.17,16.62 & 
$2.33 \pm 0.09$ & 
0.2405 & 
0.4700 & 
$299 \pm 16$ & 
LRG &
Yes\\ 

SDSS J095944.07$+$041017.0 & 
0572-52289-495 & 
18.52,17.48,17.02 & 
$1.21 \pm 0.04$ & 
0.1260 & 
0.5350 & 
$212 \pm 12$ & 
MAIN &
Yes\\ 

SDSS J102551.31$-$003517.4 & 
0272-51941-151 & 
17.07,16.03,15.57 & 
$4.05 \pm 0.08$ & 
0.1589 & 
0.2764 & 
$247 \pm 11$ & 
LRG &
?\\ 

SDSS J111739.60$+$053413.9 & 
0835-52326-571 & 
18.72,17.56,17.12 & 
$2.49 \pm 0.11$ & 
0.2285 & 
0.8232 & 
$279 \pm 21$ & 
MAIN &
?\\ 

SDSS J120540.43$+$491029.3 & 
0969-52442-134 & 
18.40,17.22,16.65 & 
$2.30 \pm 0.10$ & 
0.2150 & 
0.4808 & 
$235 \pm 10$ & 
MAIN &
Yes\\ 

SDSS J125028.25$+$052349.0 & 
0847-52426-549 & 
18.40,17.26,16.77 & 
$1.76 \pm 0.07$ & 
0.2318 & 
0.7946 & 
$254 \pm 14$ & 
MAIN &
Yes\\ 

SDSS J125135.70$-$020805.1 & 
0337-51997-480 & 
18.58,17.59,17.24 & 
$3.64 \pm 0.19$ & 
0.2243 & 
0.7843 & 
$216 \pm 23$ & 
MAIN &
Yes\\ 

SDSS J125919.05$+$613408.6 & 
0783-52325-279 & 
18.80,17.46,17.01 & 
$1.94 \pm 0.07$ & 
0.2333 & 
0.4488 & 
$263 \pm 17$ & 
LRG &
?\\ 

SDSS J133045.53$-$014841.6 & 
0910-52377-503 & 
18.34,17.45,17.05 & 
$0.84 \pm 0.04$ & 
0.0808 & 
0.7115 & 
$178 \pm 09$ & 
MAIN &
Yes\\ 

SDSS J140228.21$+$632133.5 & 
0605-52353-503 & 
18.26,16.98,16.49 & 
$2.67 \pm 0.08$ & 
0.2046 & 
0.4814 & 
$275 \pm 15$ & 
LRG &
Yes\\ 

SDSS J142015.85$+$601914.8 & 
0788-52338-605 & 
16.39,15.56,15.17 & 
$2.17 \pm 0.03$ & 
0.0629 & 
0.5350 & 
$194 \pm 05$ & 
MAIN &
Yes\\ 

SDSS J154731.22$+$572000.0 & 
0617-52072-561 & 
17.94,16.84,16.39 & 
$2.56 \pm 0.06$ & 
0.1883 & 
0.3955 & 
$243 \pm 11$ & 
LRG &
?\\ 

SDSS J161843.10$+$435327.4 & 
0815-52374-337 & 
18.78,17.60,17.09 & 
$1.34 \pm 0.05$ & 
0.1989 & 
0.6656 & 
$257 \pm 25$ & 
MAIN &
Yes\\ 

SDSS J162746.44$-$005357.5 & 
0364-52000-084 & 
18.54,17.29,16.89 & 
$2.08 \pm 0.08$ & 
0.2076 & 
0.5241 & 
$275 \pm 12$ & 
LRG &
Yes\\ 

SDSS J163028.15$+$452036.2 & 
0626-52057-518 & 
18.84,17.41,16.92 & 
$2.02 \pm 0.07$ & 
0.2479 & 
0.7933 & 
$260 \pm 16$ & 
LRG &
Yes\\ 

SDSS J163602.61$+$470729.5 & 
0627-52144-464 & 
18.92,17.68,17.18 & 
$1.48 \pm 0.05$ & 
0.2282 & 
0.6745 & 
$221 \pm 15$ & 
MAIN &
?\\ 

SDSS J170216.76$+$332044.7 & 
0973-52426-464 & 
18.01,16.91,16.40 & 
$2.80 \pm 0.07$ & 
0.1784 & 
0.4357 & 
$239 \pm 14$ & 
LRG &
?\\ 

SDSS J171837.39$+$642452.2 & 
0352-51789-563 & 
16.86,15.97,15.54 & 
$3.67 \pm 0.07$ & 
0.0899 & 
0.7367 & 
$270 \pm 16$ & 
MAIN &
Yes\\ 

SDSS J230053.14$+$002237.9 & 
0677-52606-520 & 
18.97,17.63,17.14 & 
$1.76 \pm 0.10$ & 
0.2285 & 
0.4635 & 
$283 \pm 18$ & 
LRG &
Yes\\ 

SDSS J230321.72$+$142217.9 & 
0743-52262-304 & 
17.58,16.39,15.96 & 
$3.02 \pm 0.09$ & 
0.1553 & 
0.5170 & 
$260 \pm 15$ & 
LRG &
Yes\\ 

SDSS J232120.93$-$093910.2 & 
0645-52203-517 & 
16.07,15.21,14.82 & 
$3.92 \pm 0.05$ & 
0.0819 & 
0.5324 & 
$236 \pm 07$ & 
MAIN &
Yes\\ 

SDSS J234728.08$-$000521.2 & 
0684-52523-311 & 
19.81,18.51,17.95 & 
$1.78 \pm 0.28$ & 
0.4168 & 
0.7145 & 
$330 \pm 50$ & 
LRG & 
?

\tablecomments{System Name gives truncated J2000 RA and Dec in the
format HHMMSS.ss$\pm$DDMMSS.s. De Vaucouleurs model SDSS (AB) magnitudes
have been de-reddened using dust maps from \citet*{sfd_dust}, and have
statistical errors of appproximately 0.01. De Vaucouleurs effective
radii are quoted at the intermediate axis. Parent Sample column
indicates whether candidate was selected from the SDSS luminous red
galaxy sample or from the SDSS main galaxy sample (see
\S~\ref{sec:selection}).} \enddata
\label{galtab}
\end{deluxetable*}

\begin{figure*}[t]
\centerline{\plottwo{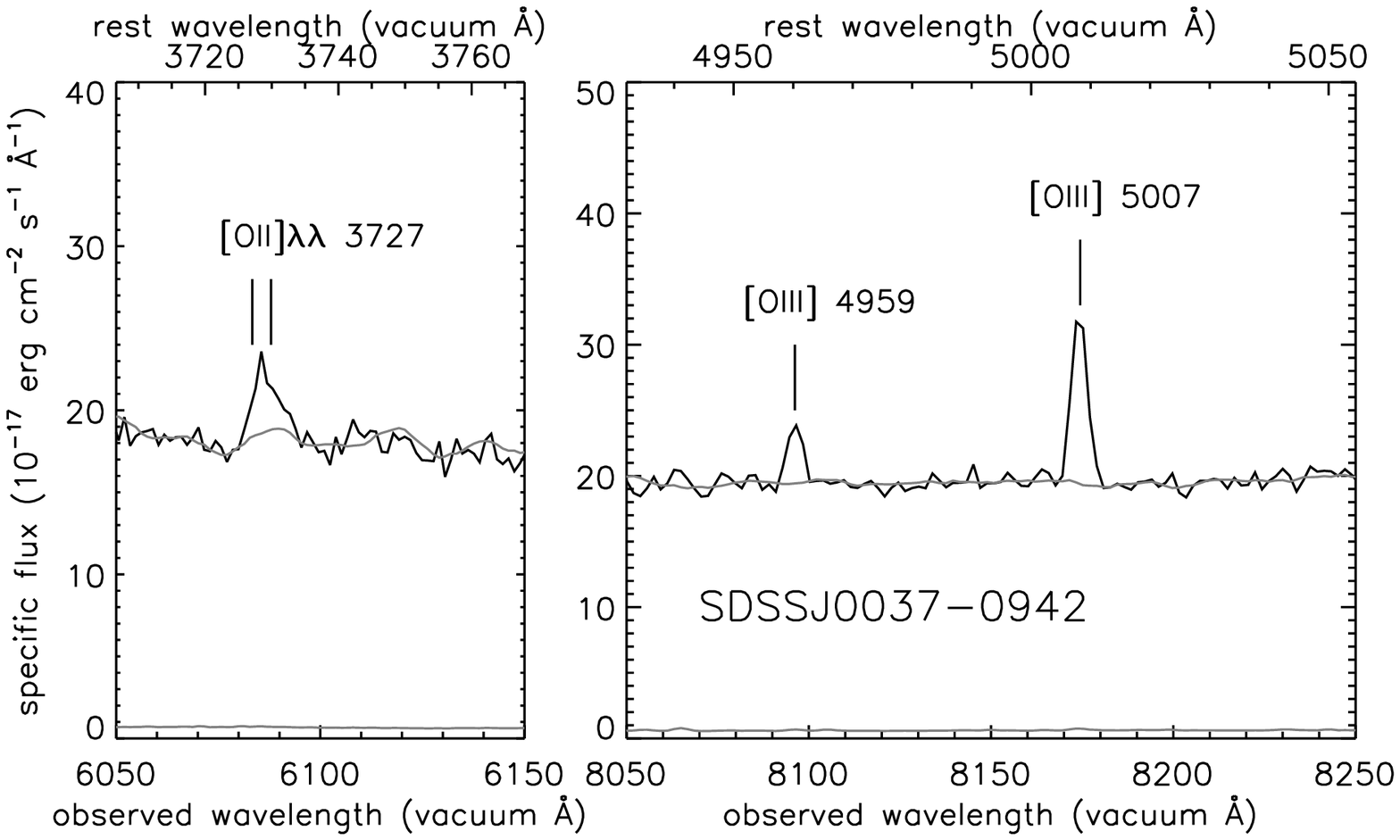}{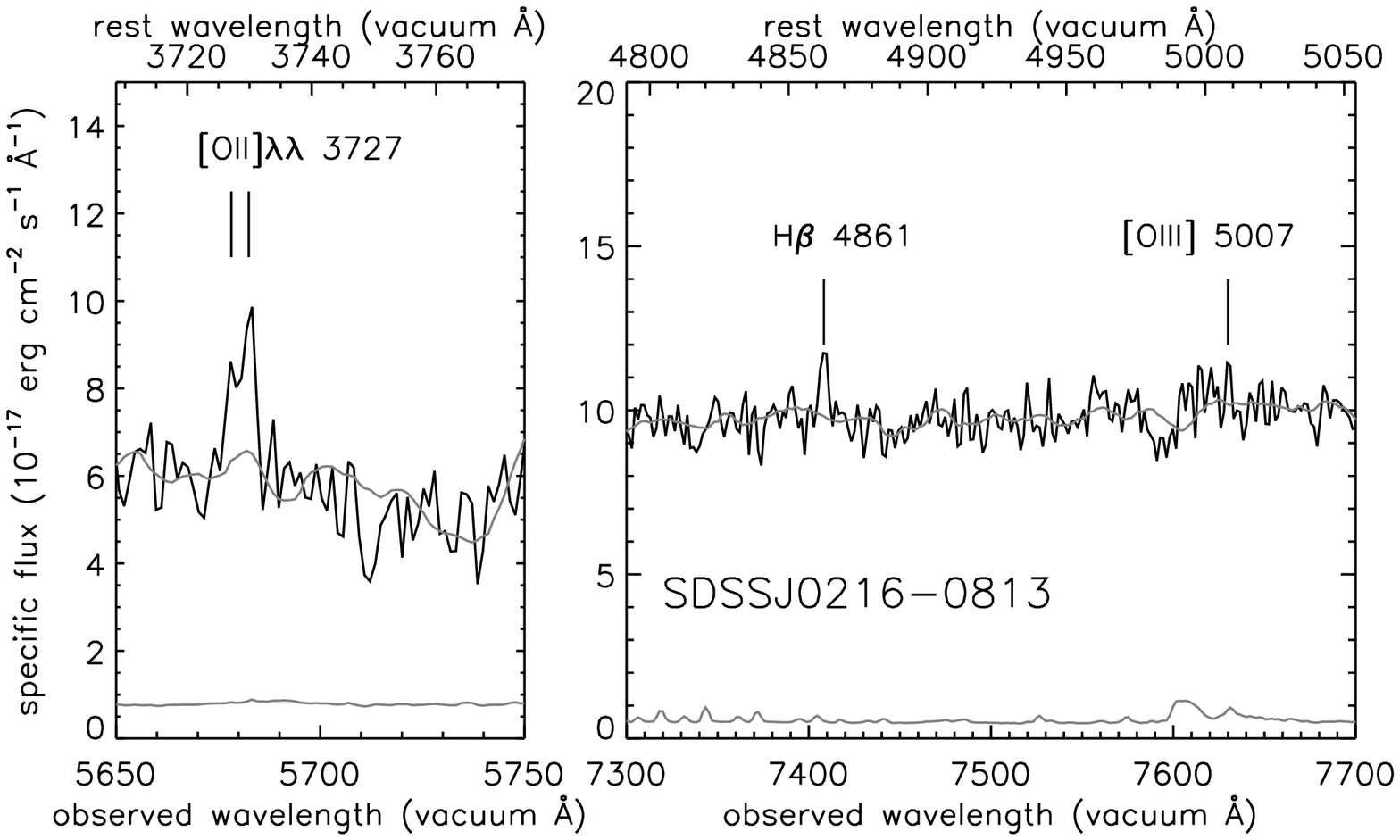}}
\centerline{\plottwo{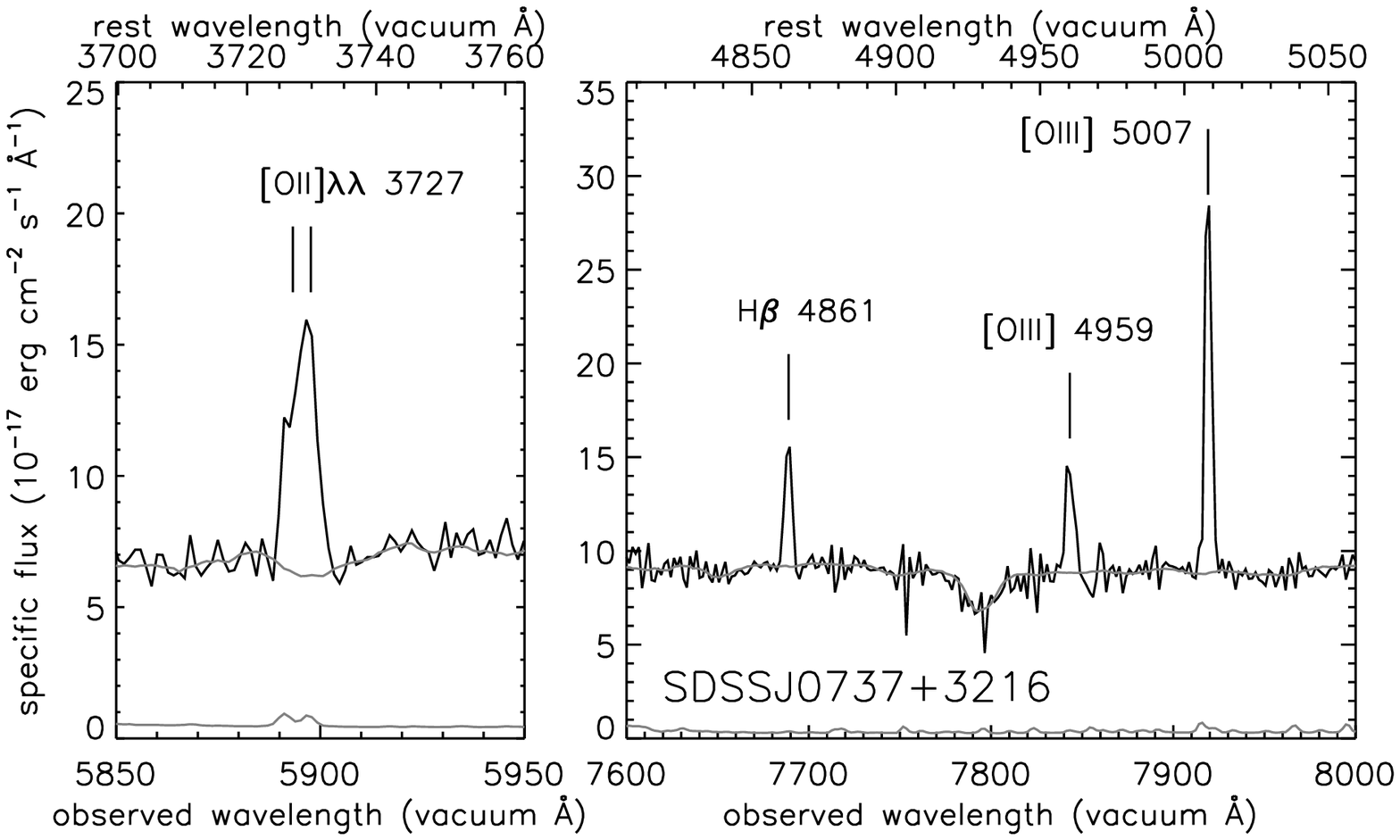}{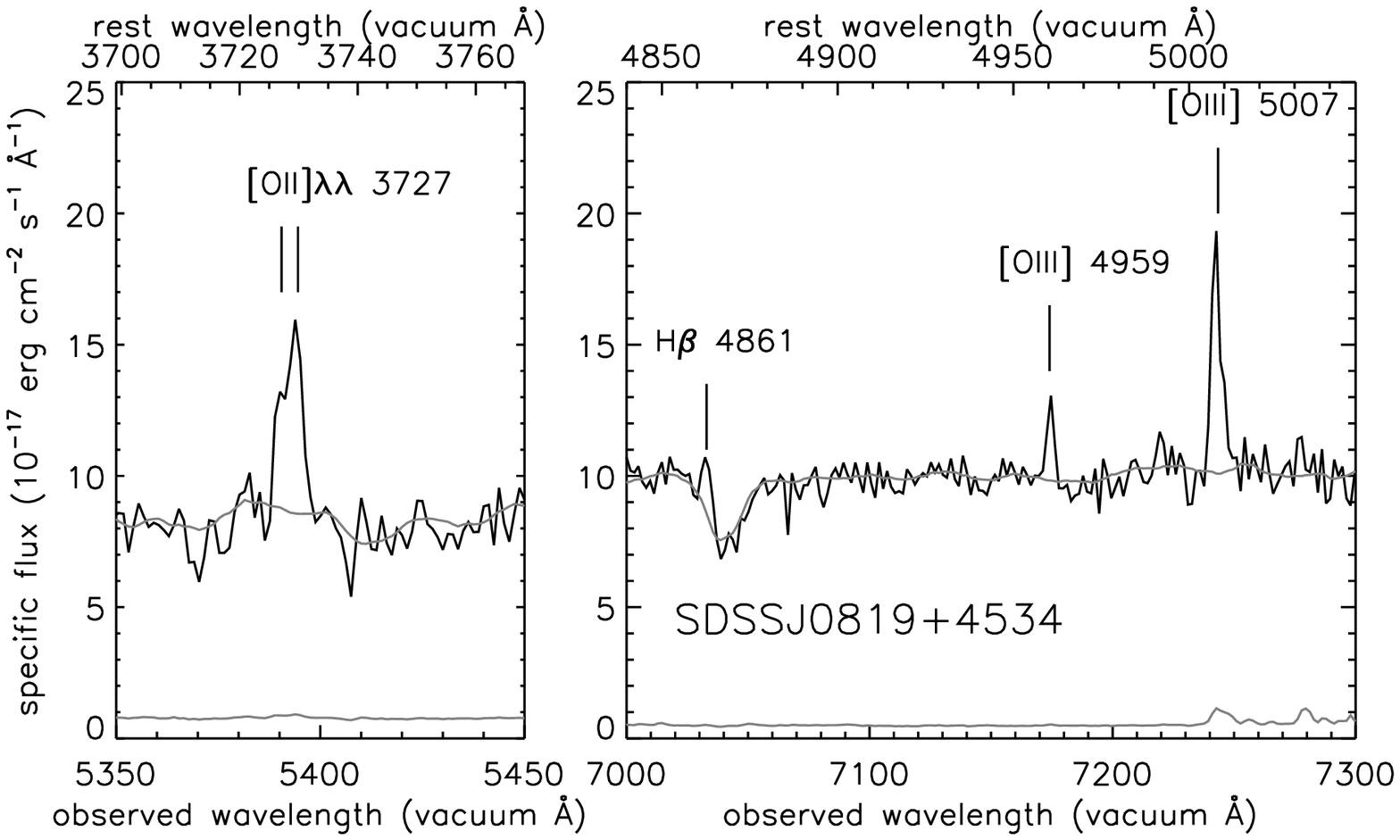}}
\centerline{\plottwo{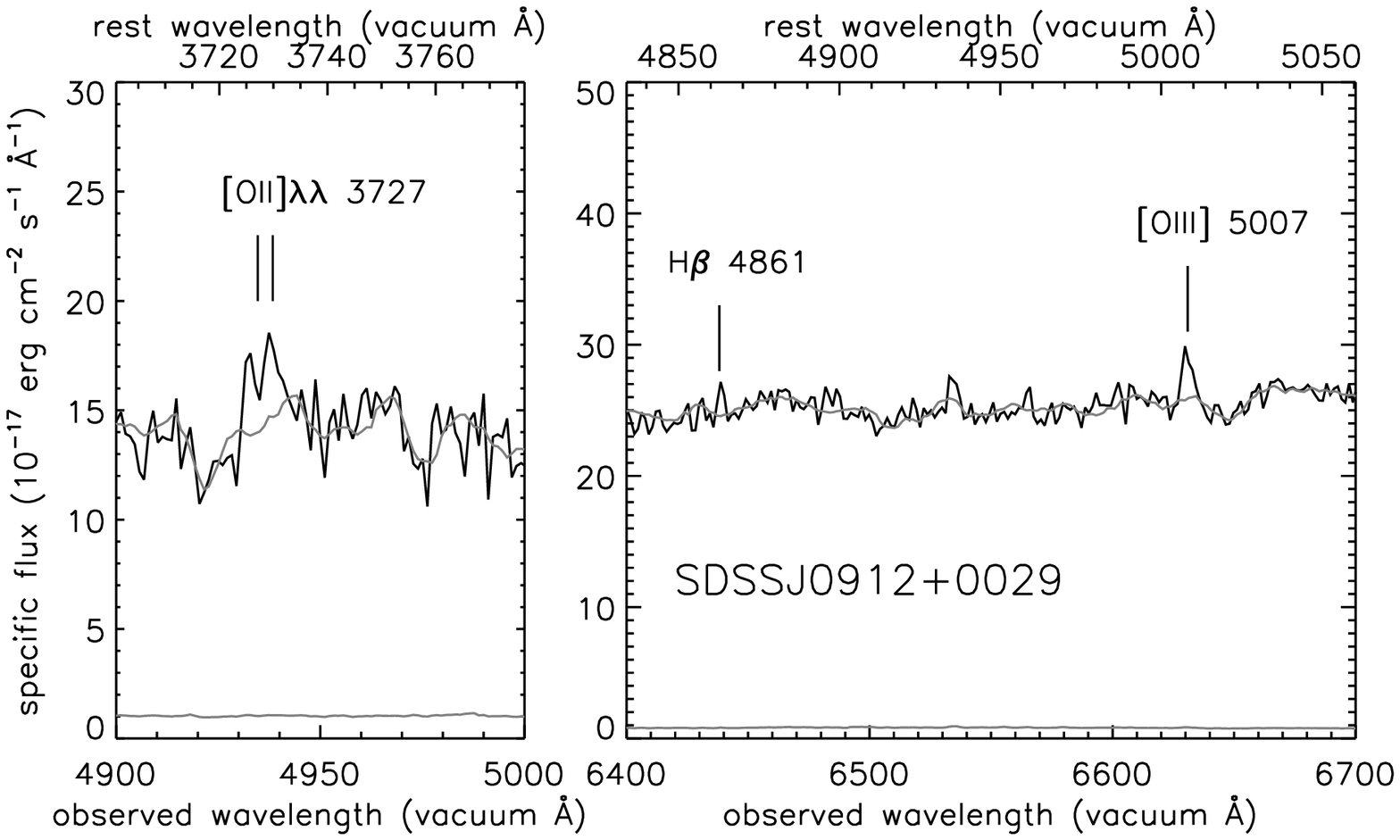}{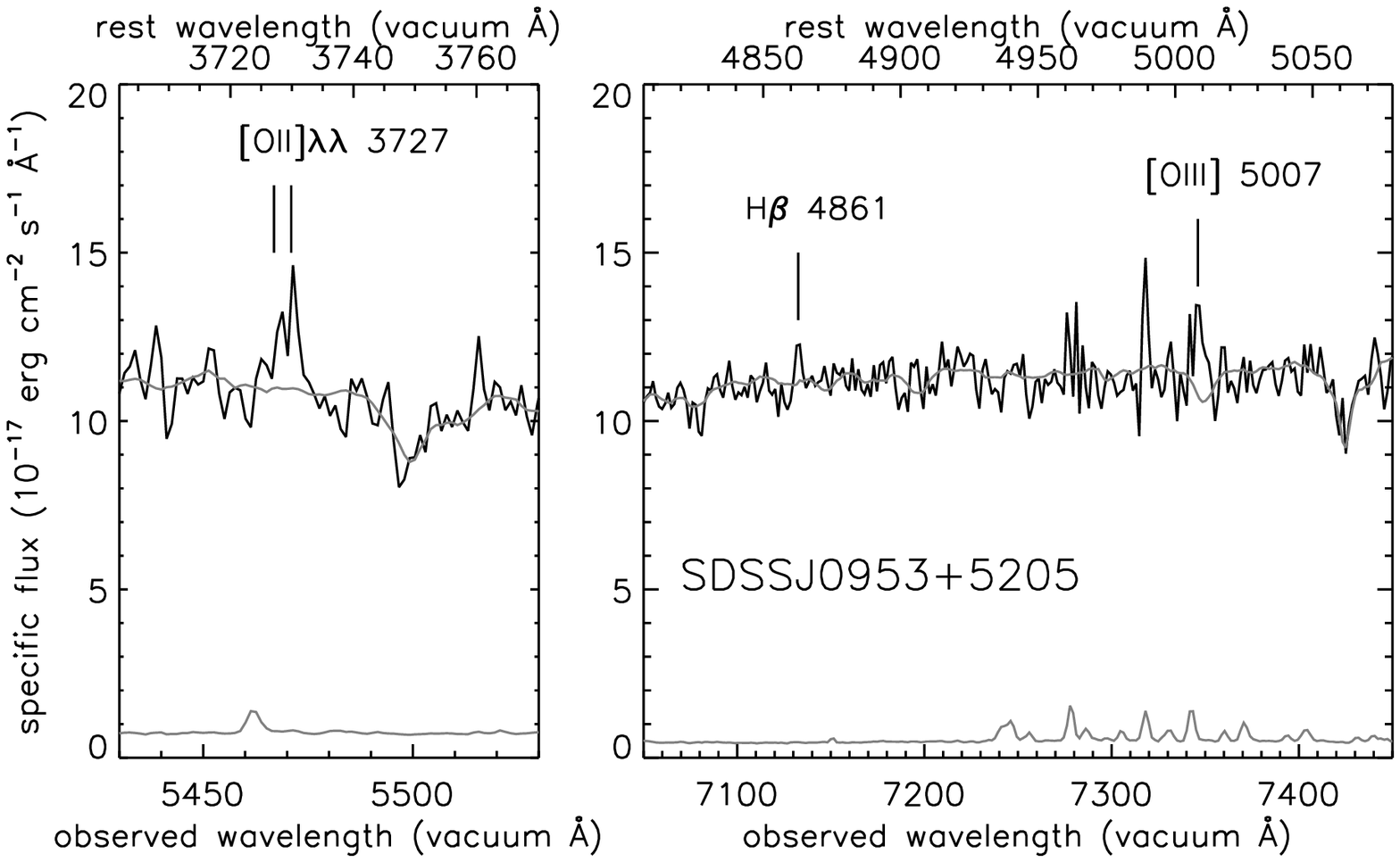}}
\centerline{\plottwo{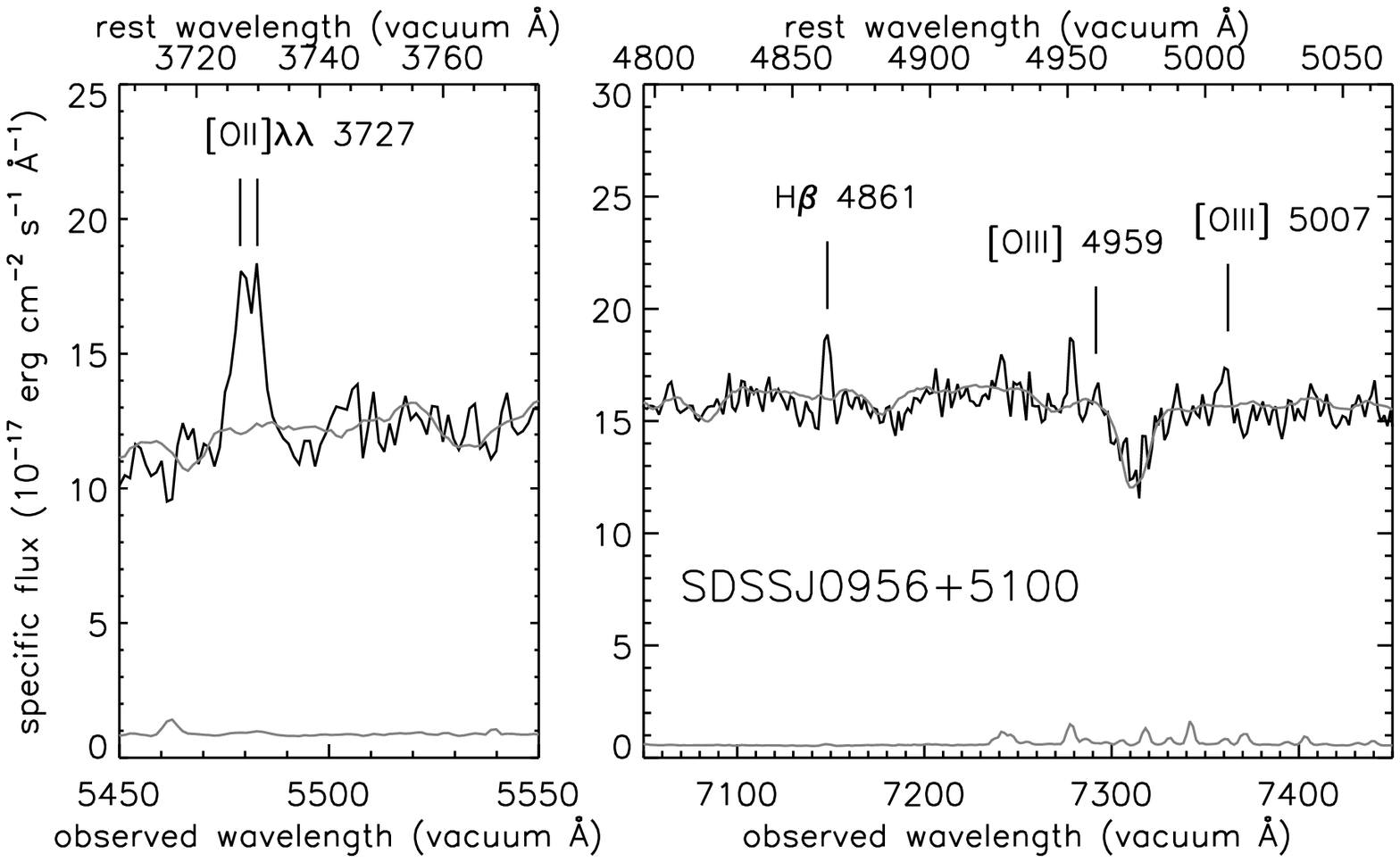}{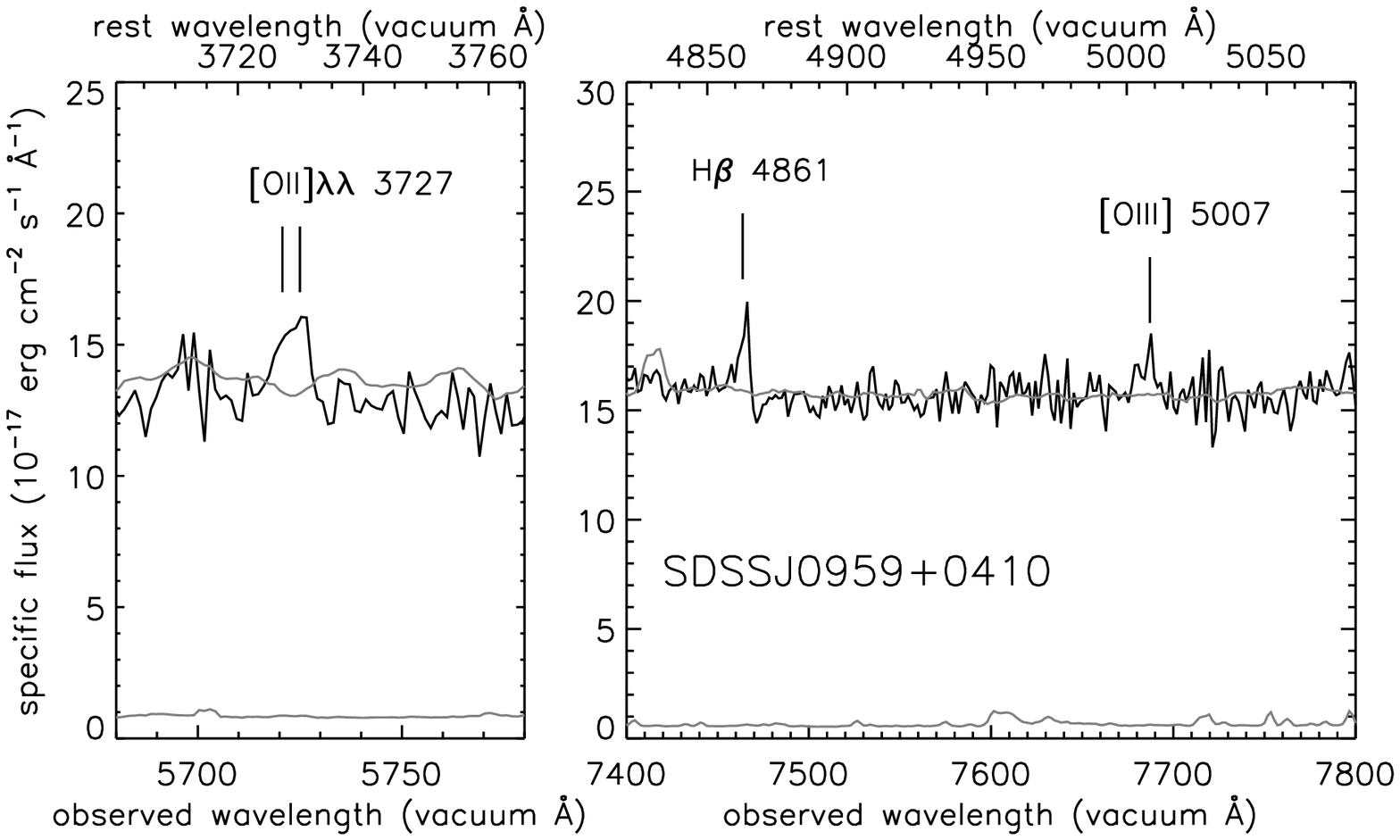}}
\caption{\label{specfig}
SDSS spectroscopy showing background line emission in SLACS target
galaxies.  Upper gray line shows SDSS template fitted to the continuum
of the foreground galaxy, and lower gray lines shows the
1-$\sigma$ noise level.}
\end{figure*}

\addtocounter{figure}{-1}

\begin{figure*}[t]
\centerline{\plottwo{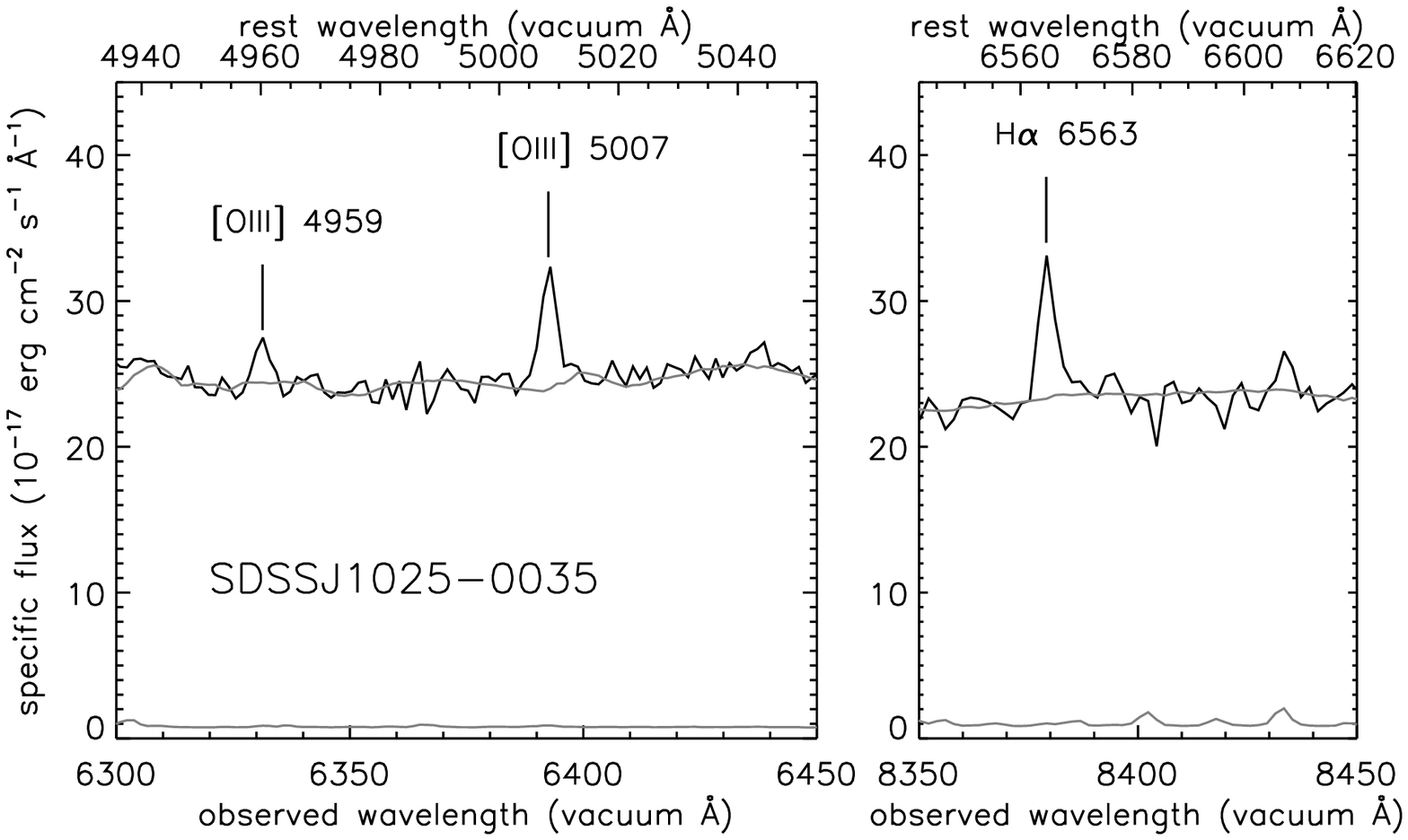}{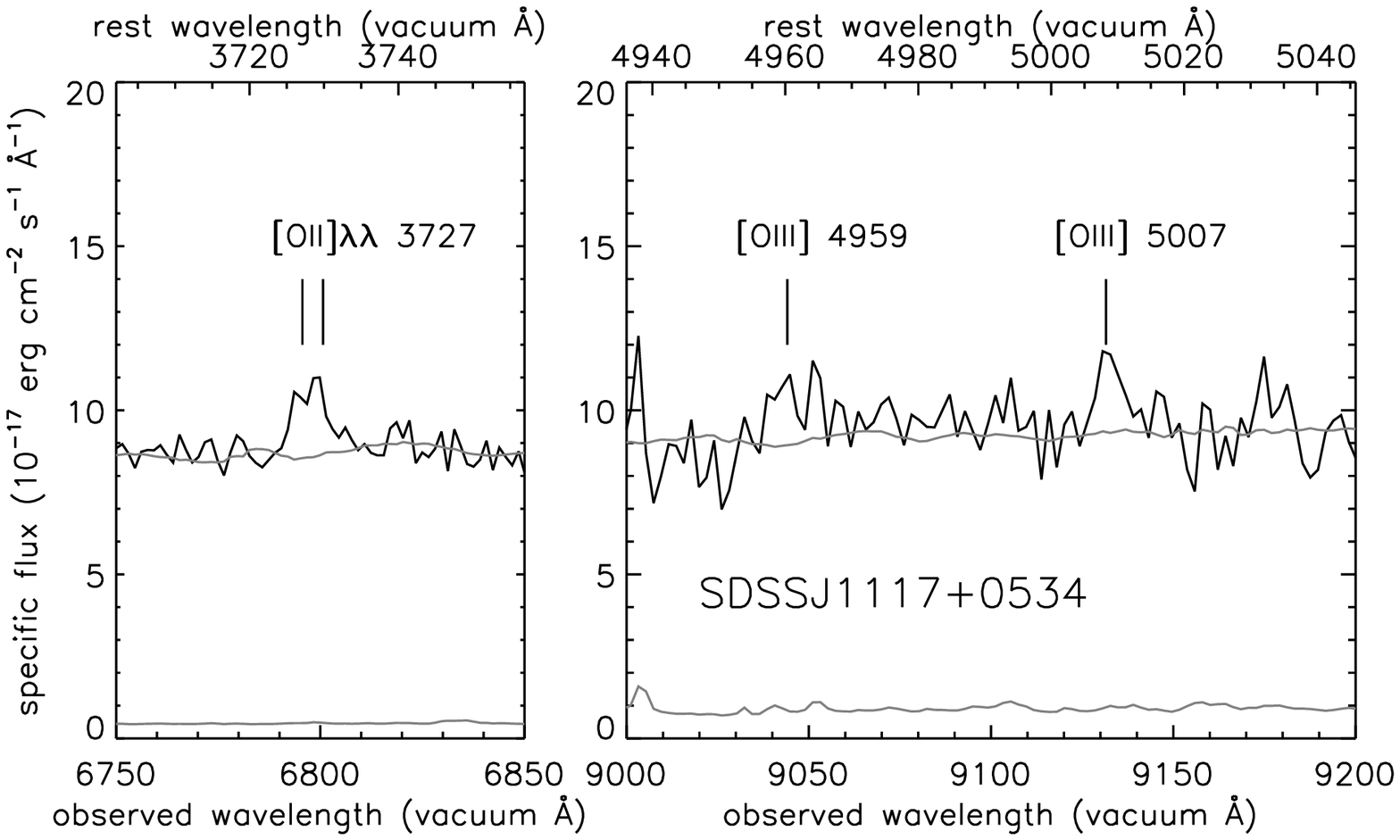}}
\centerline{\plottwo{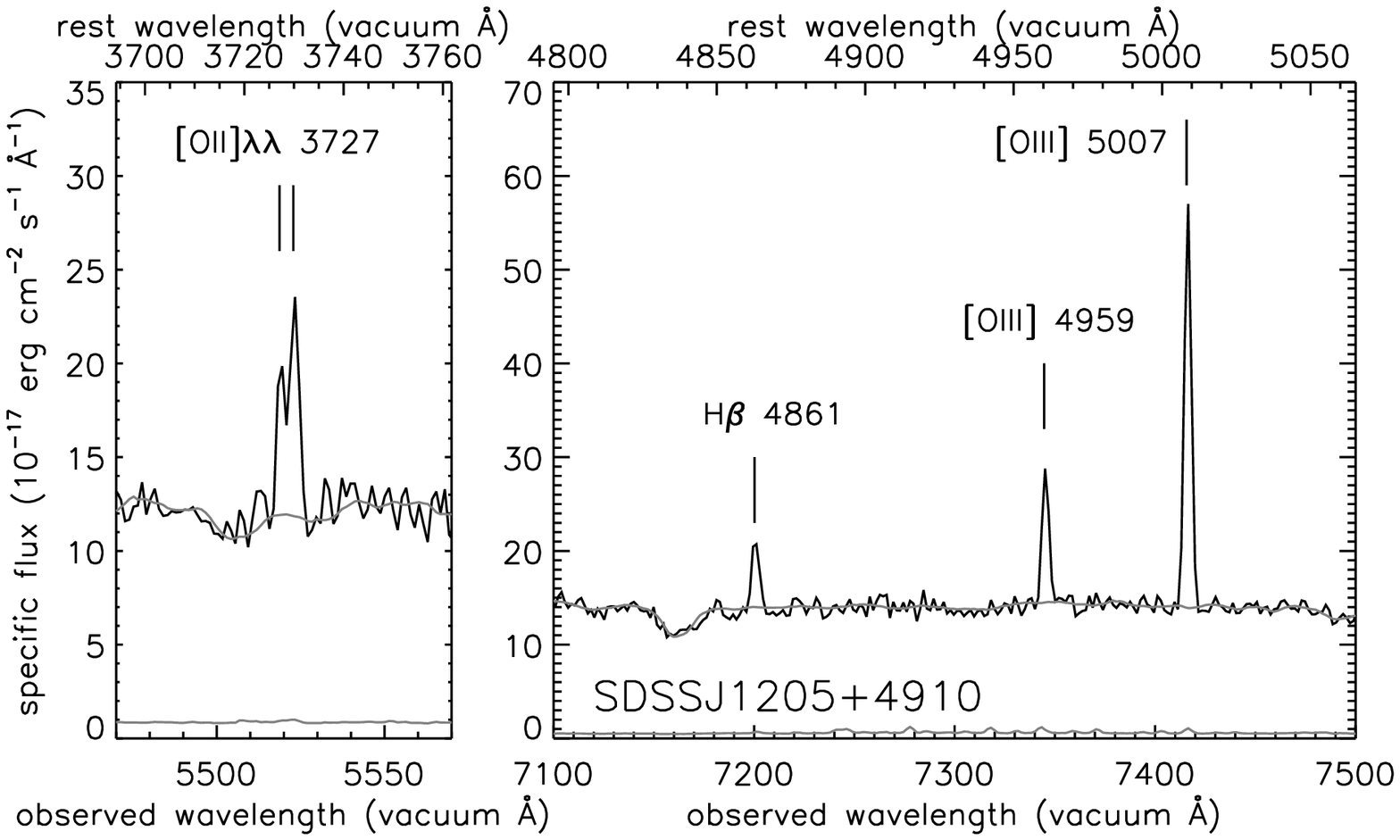}{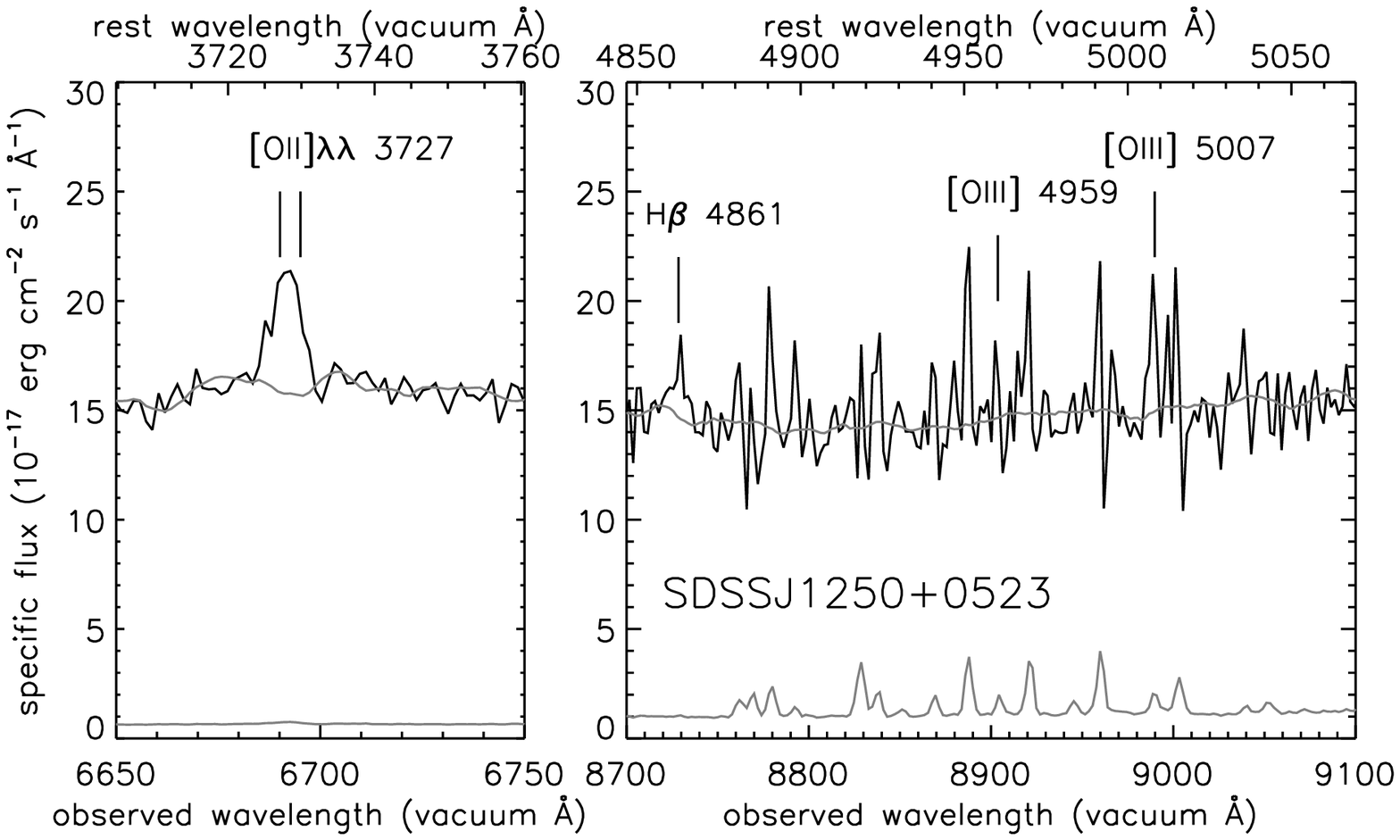}}
\centerline{\plottwo{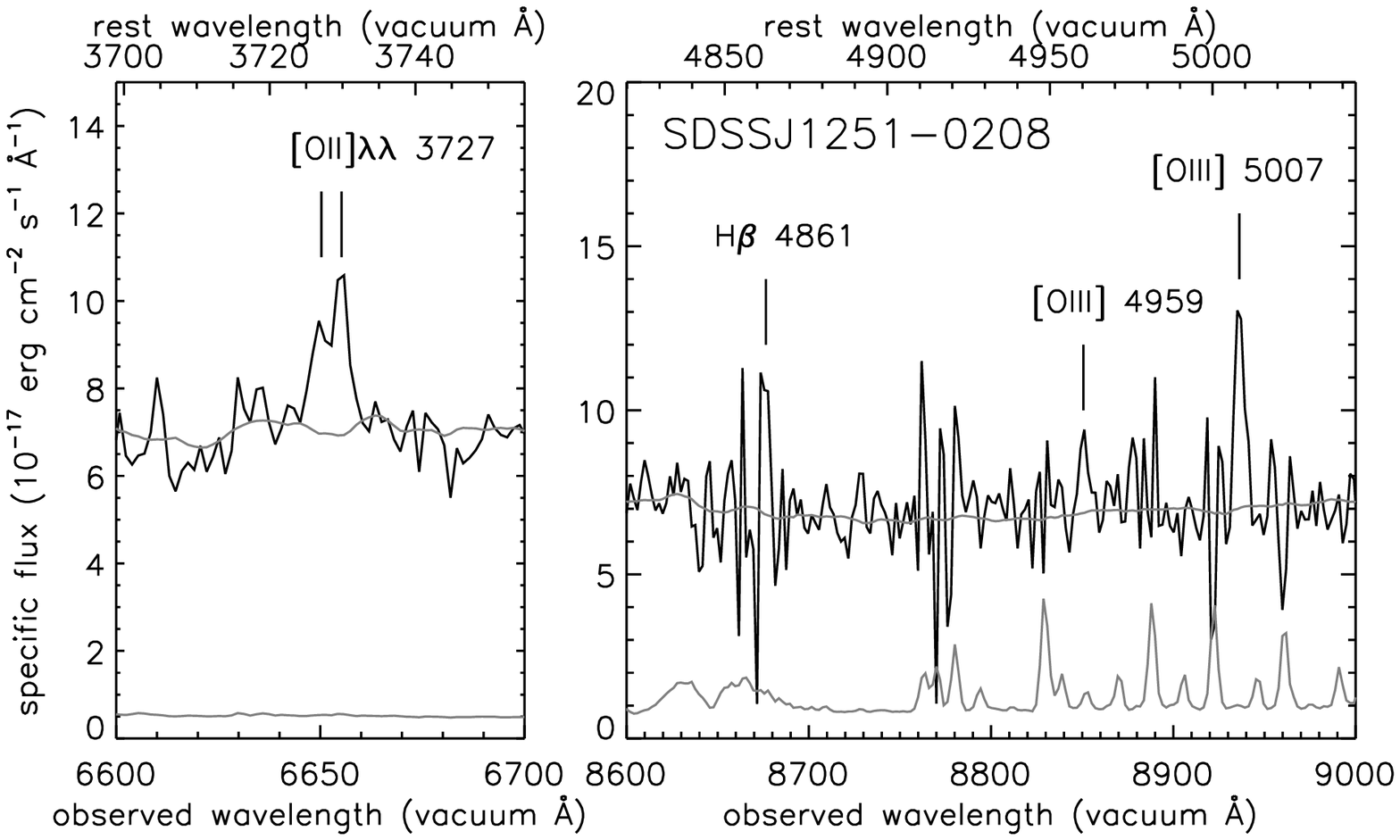}{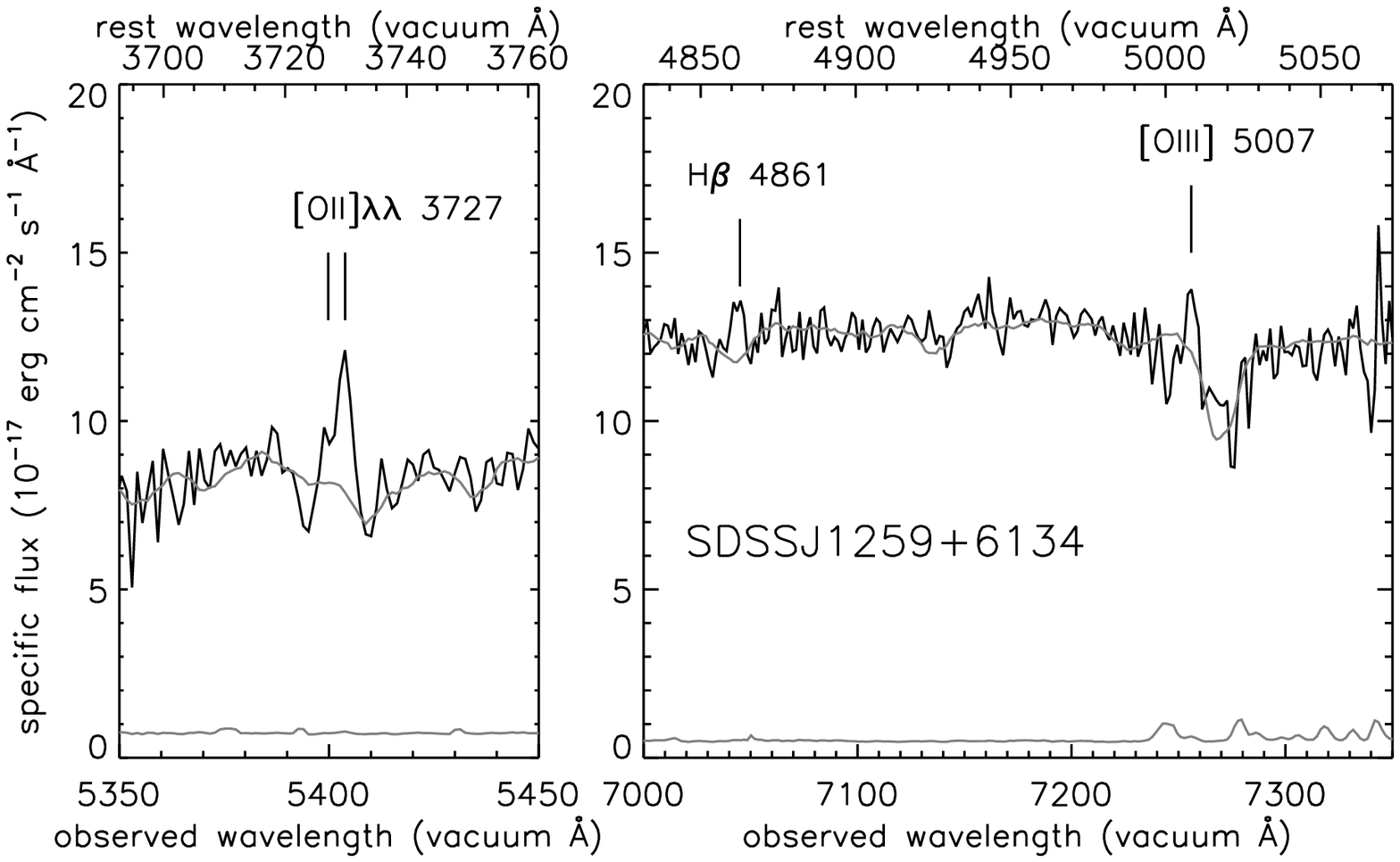}}
\centerline{\plottwo{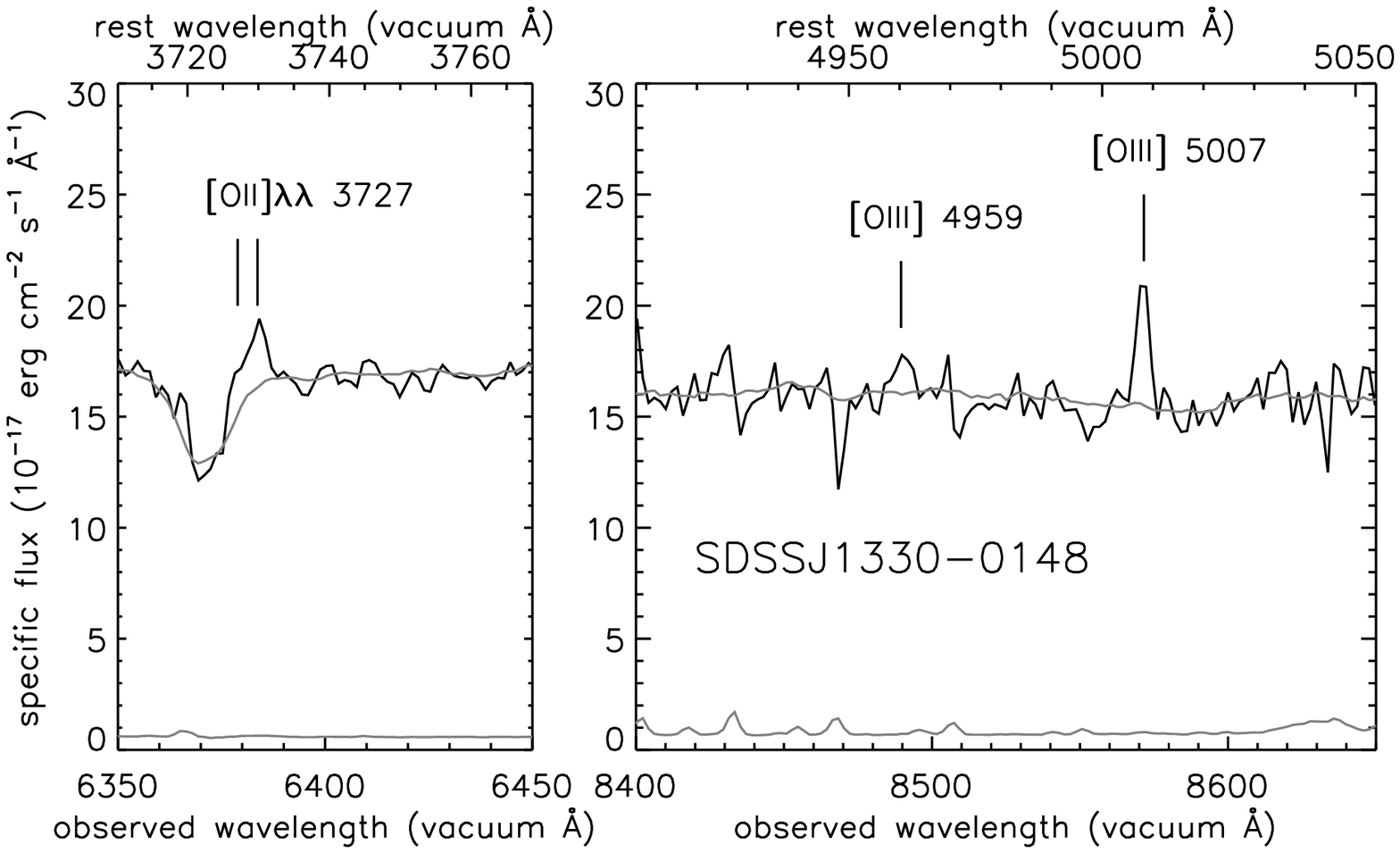}{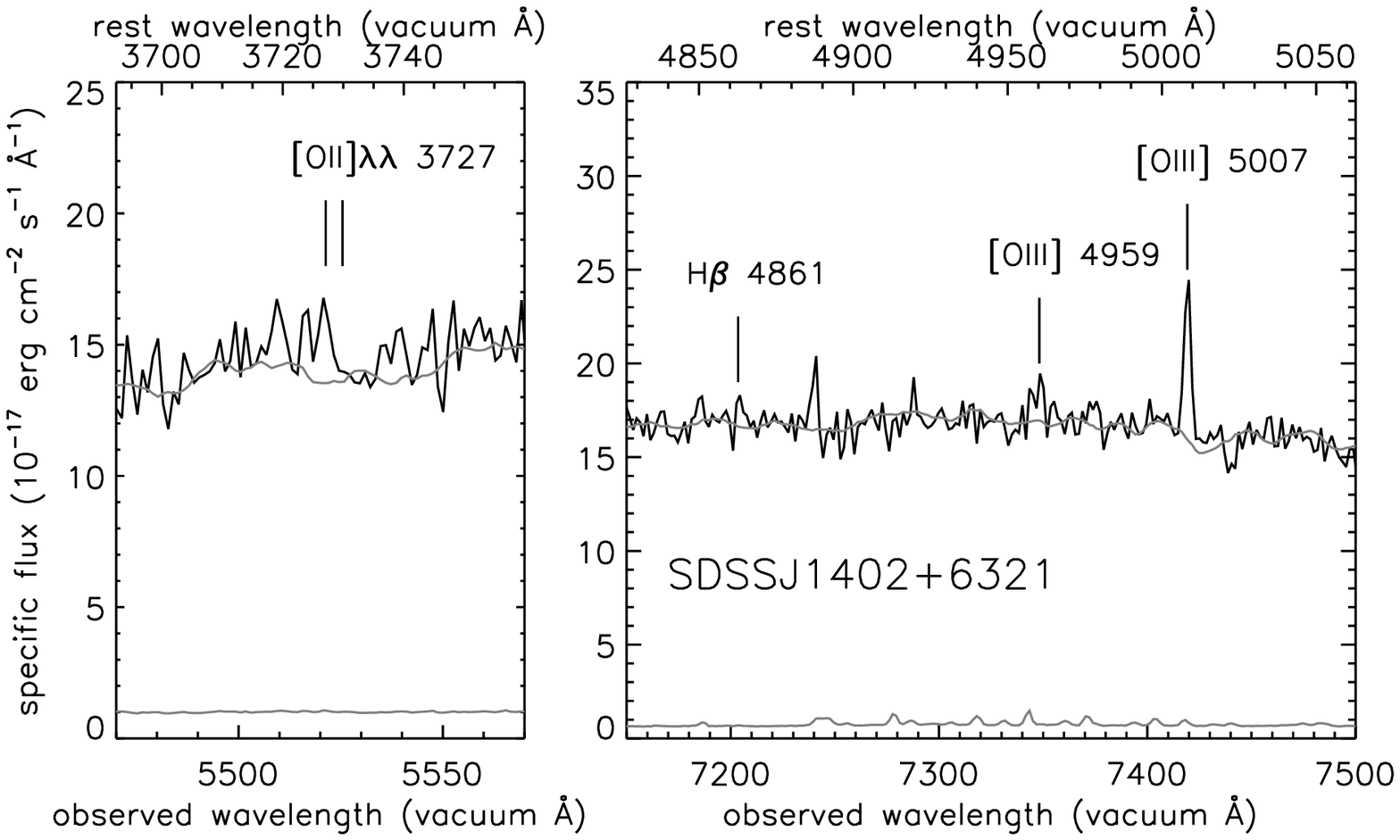}}
\centerline{\plottwo{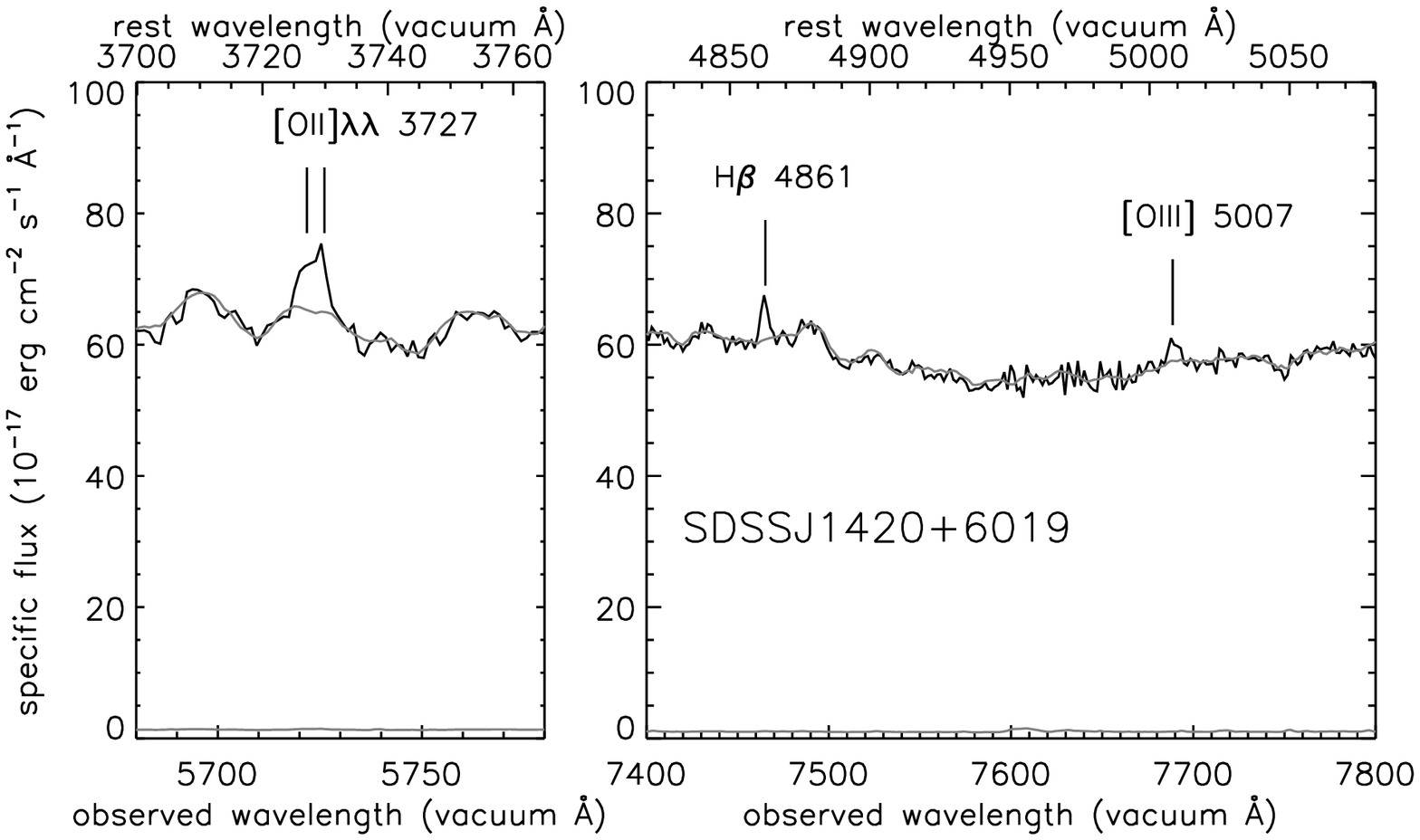}{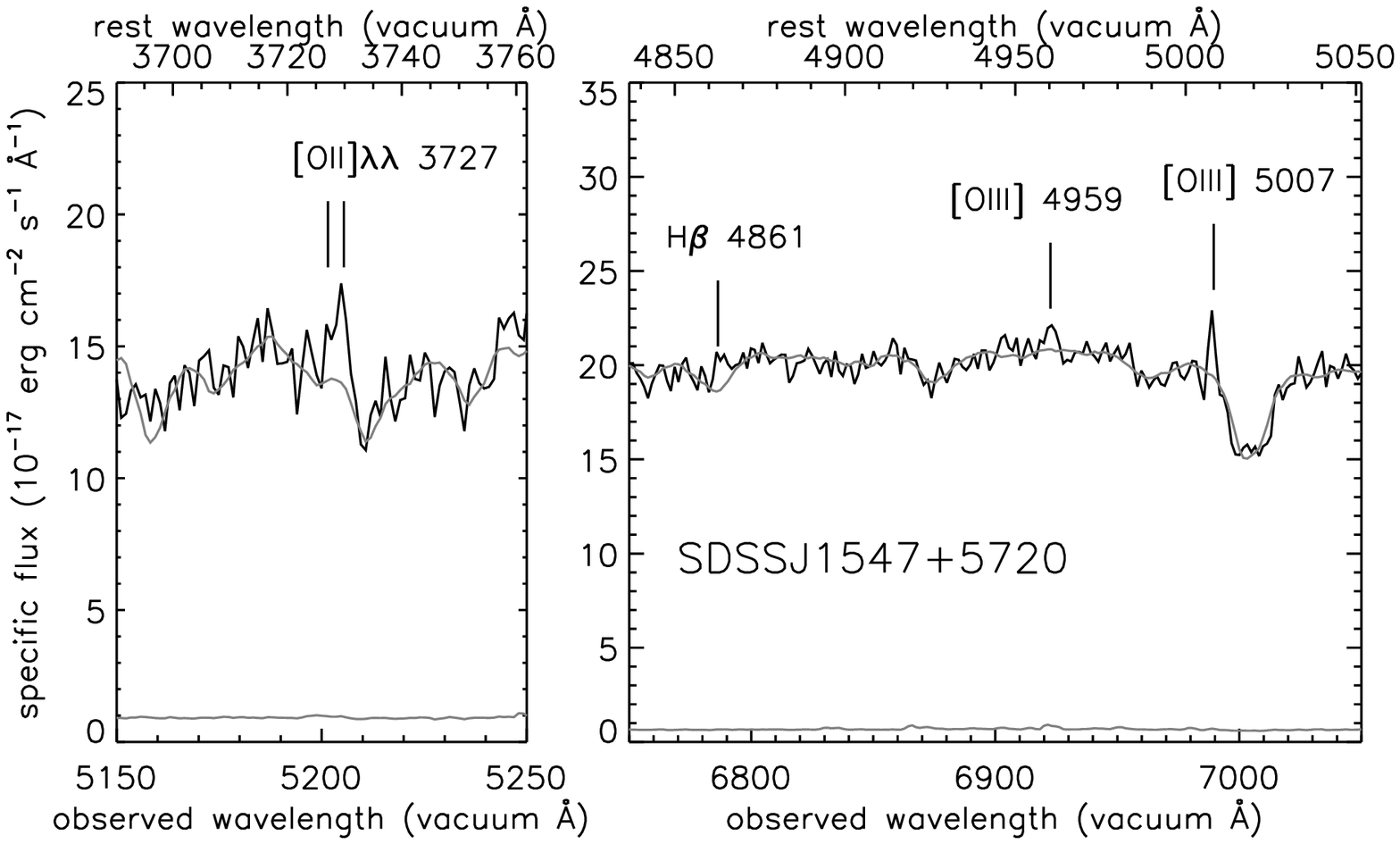}}
\caption{(continued)}
\end{figure*}

\addtocounter{figure}{-1}

\begin{figure*}[t]
\centerline{\plottwo{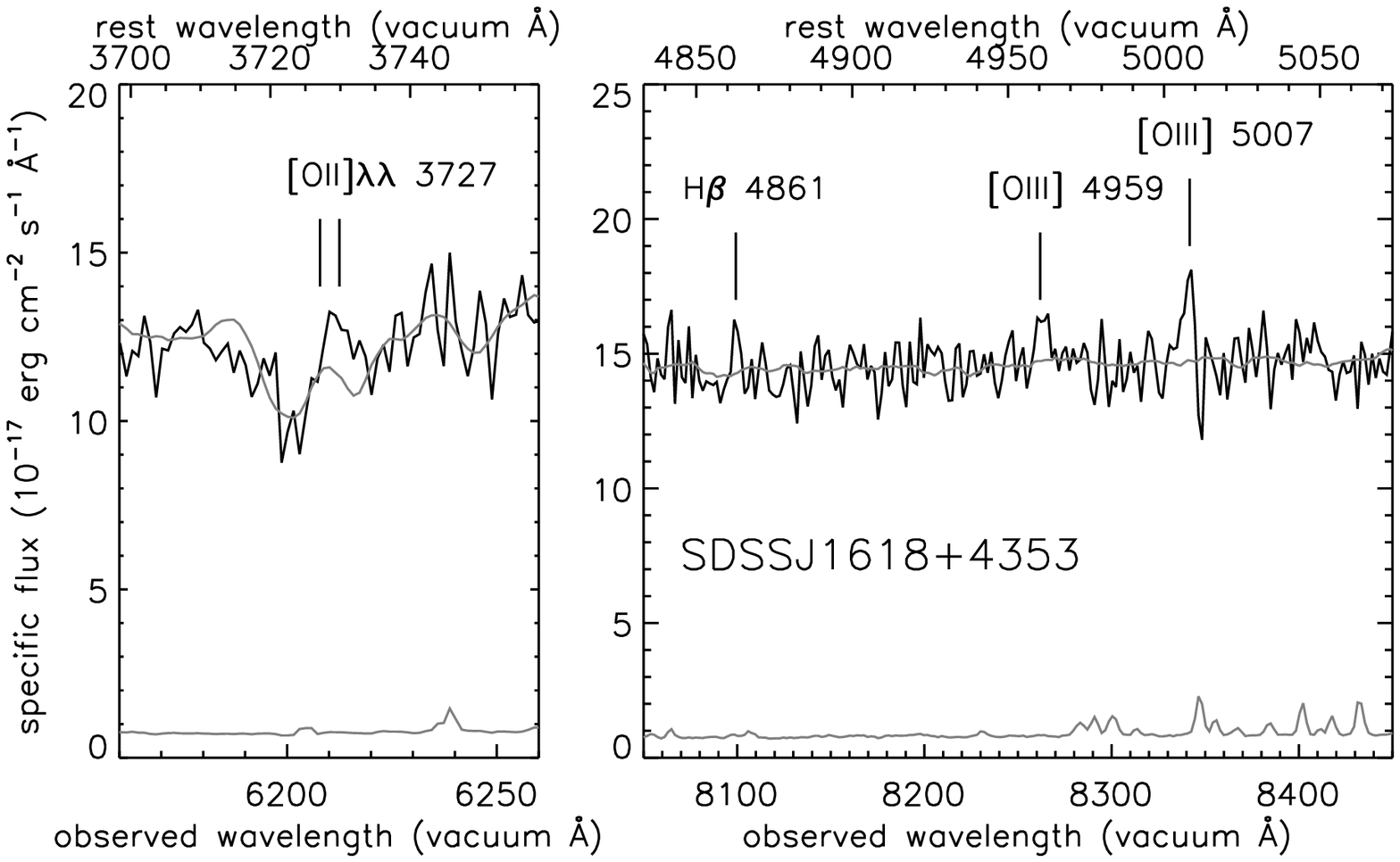}{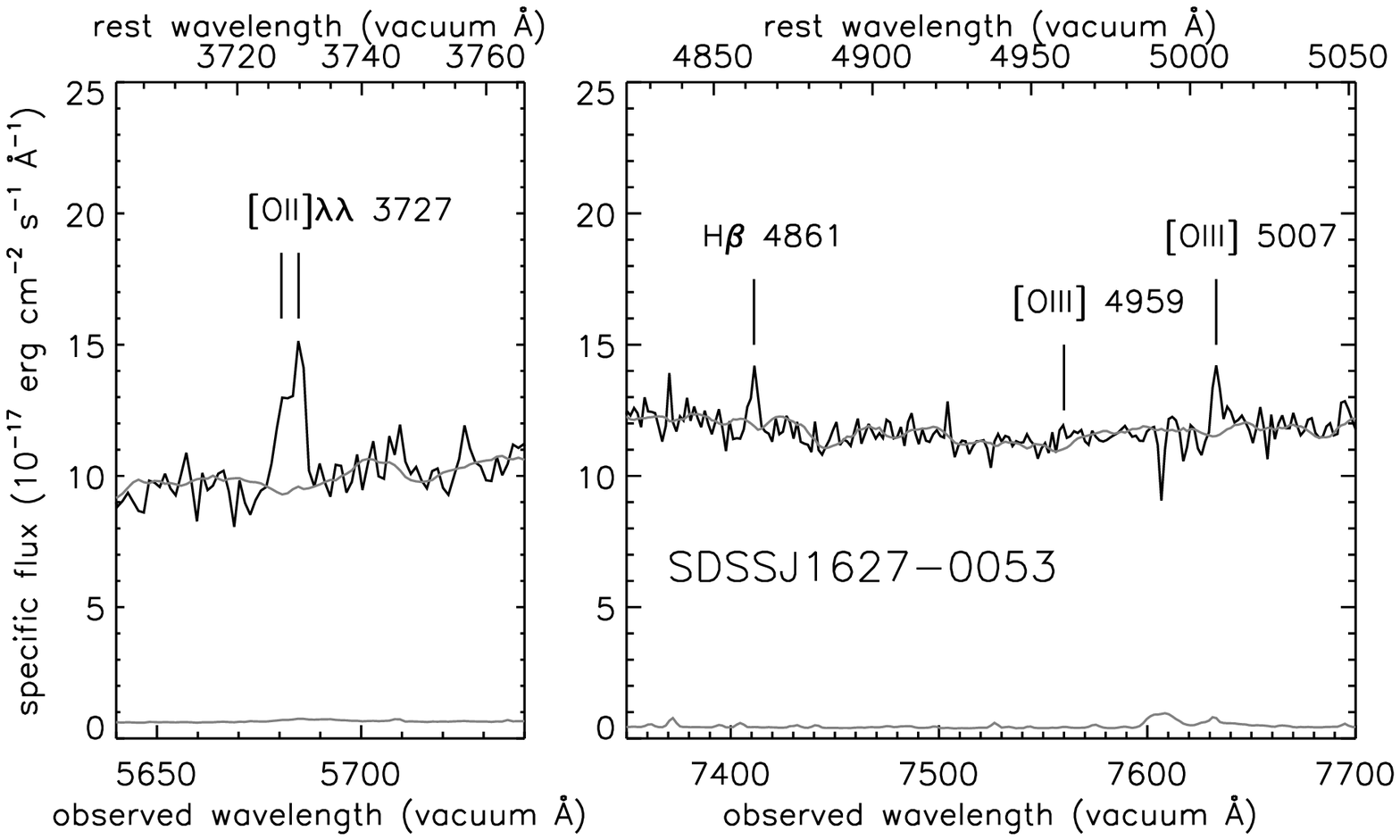}}
\centerline{\plottwo{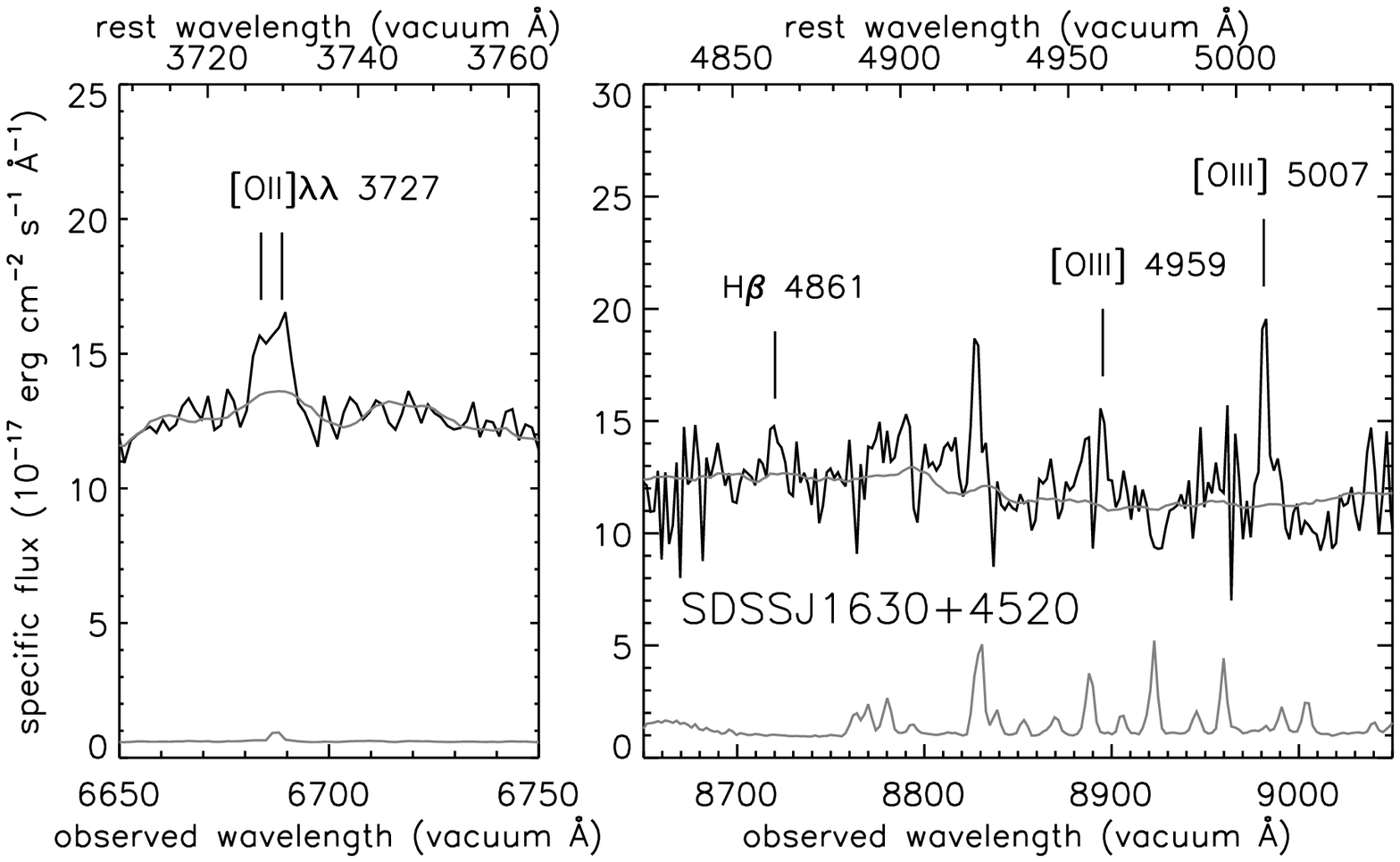}{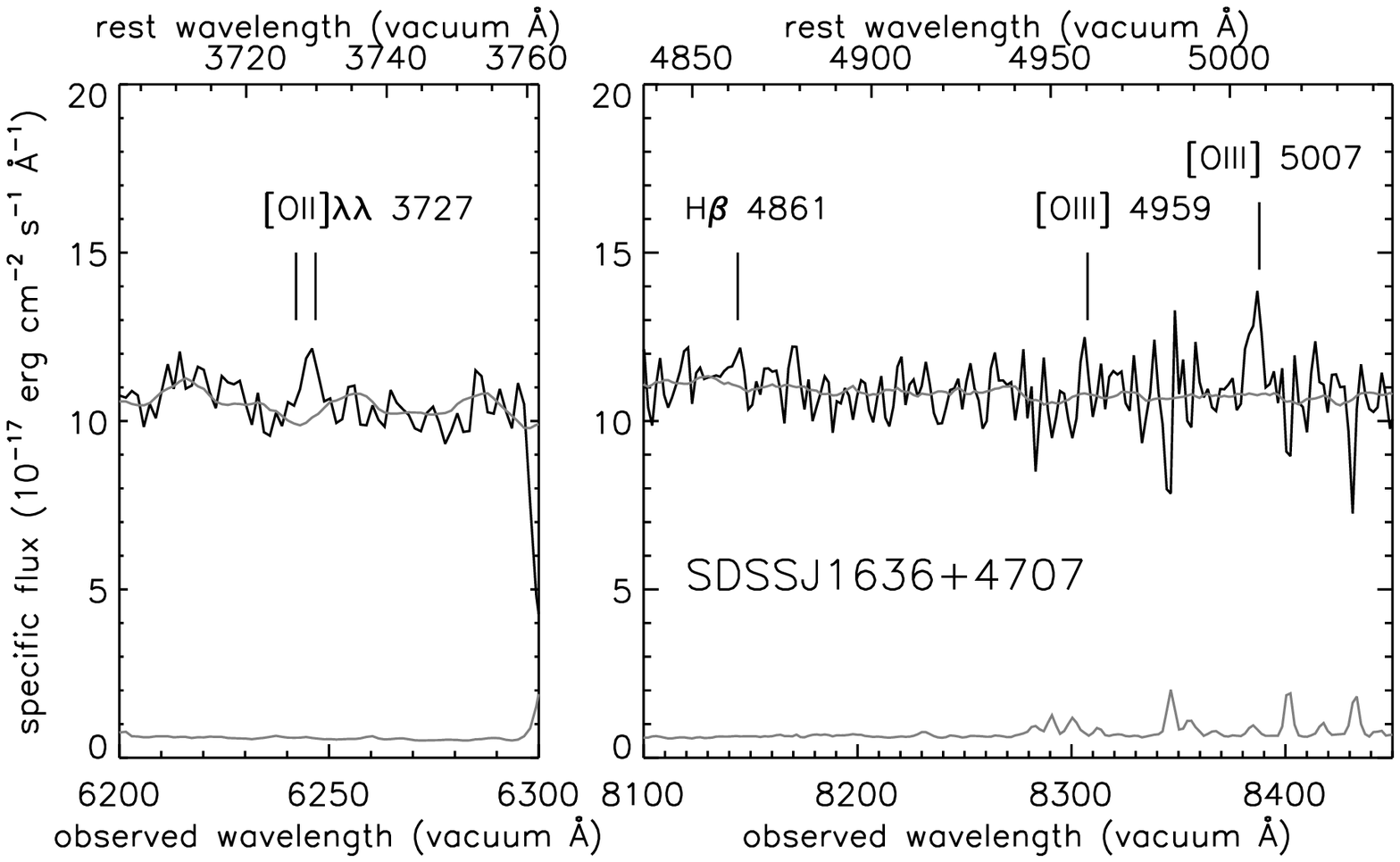}}
\centerline{\plottwo{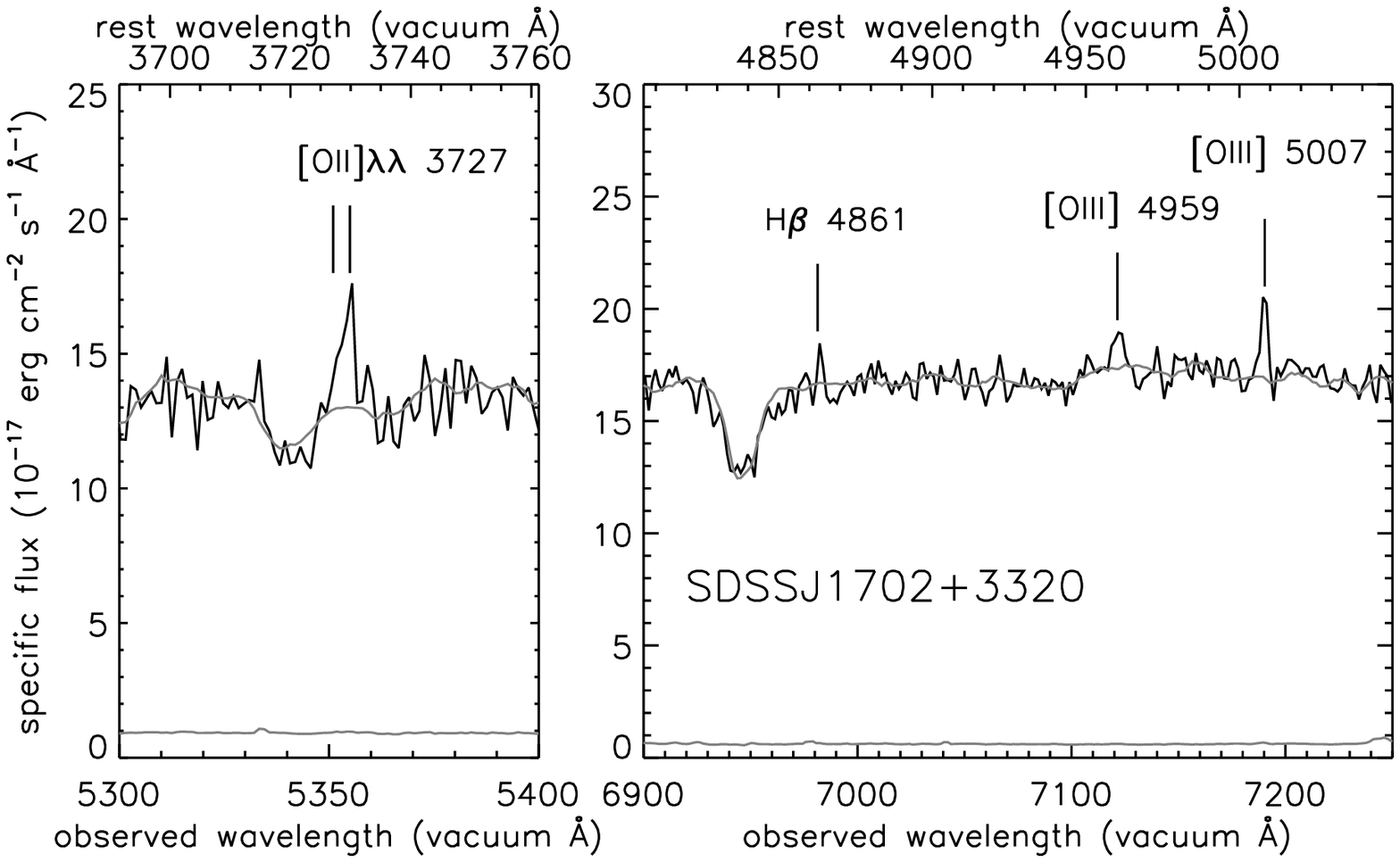}{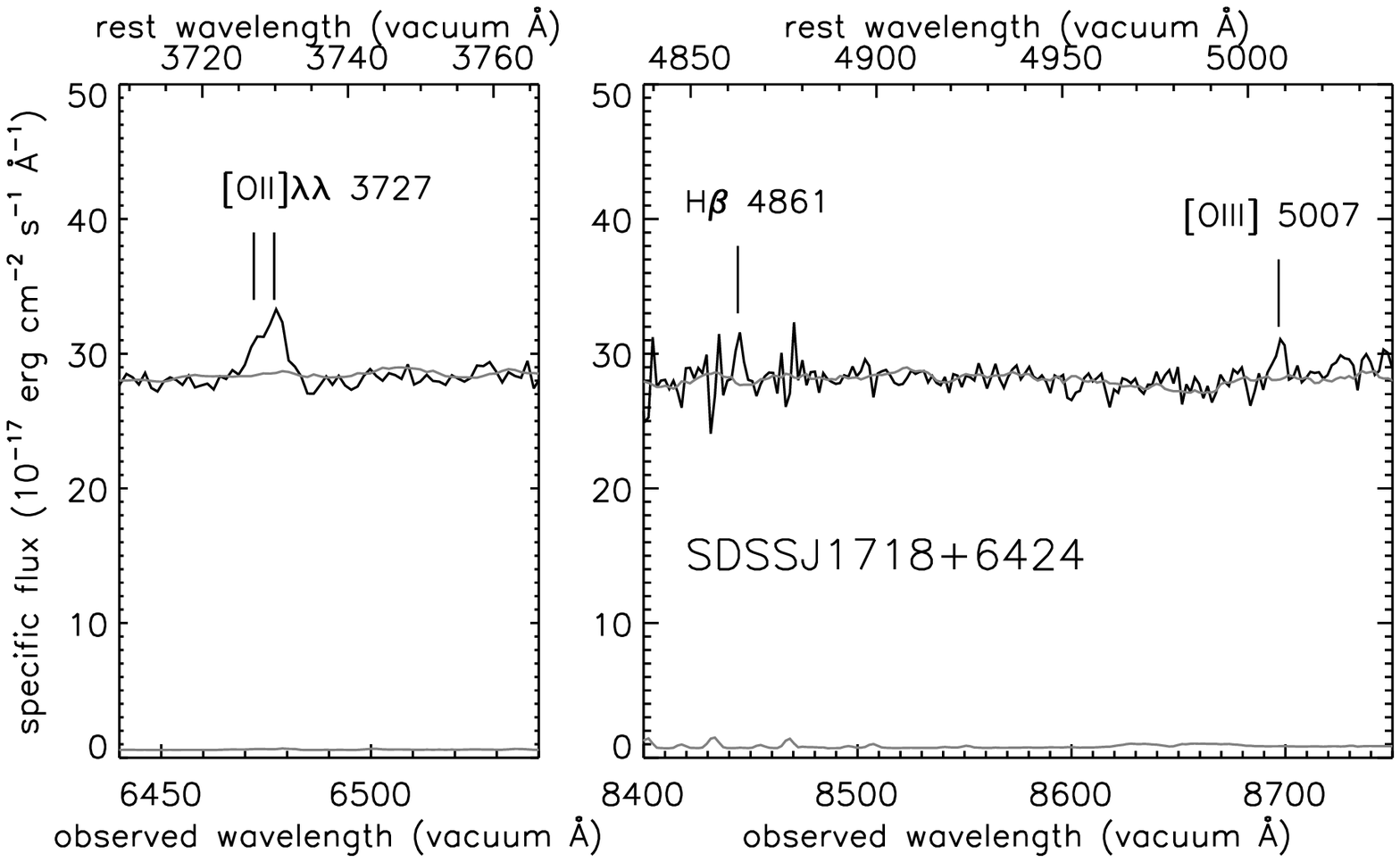}}
\centerline{\plottwo{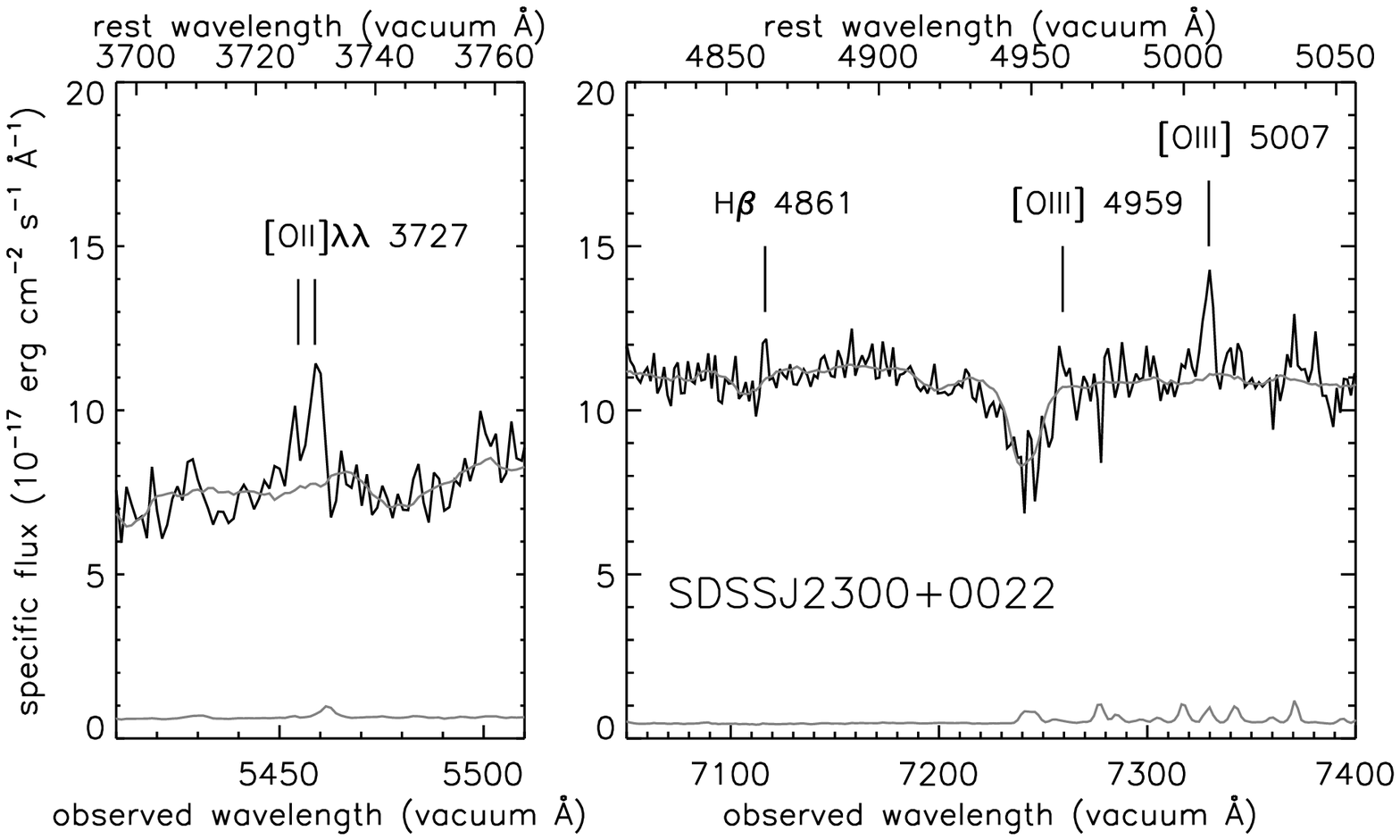}{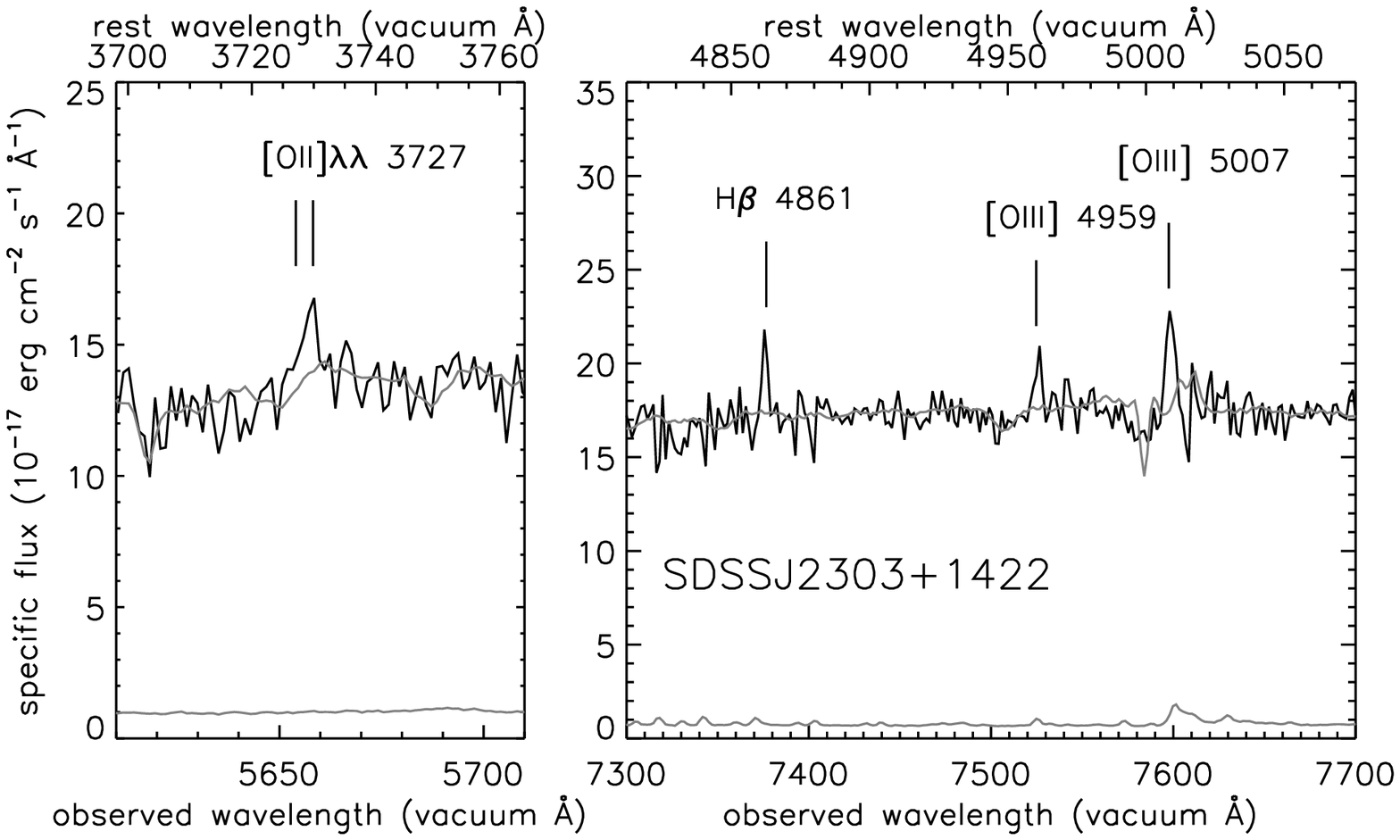}}
\centerline{\plottwo{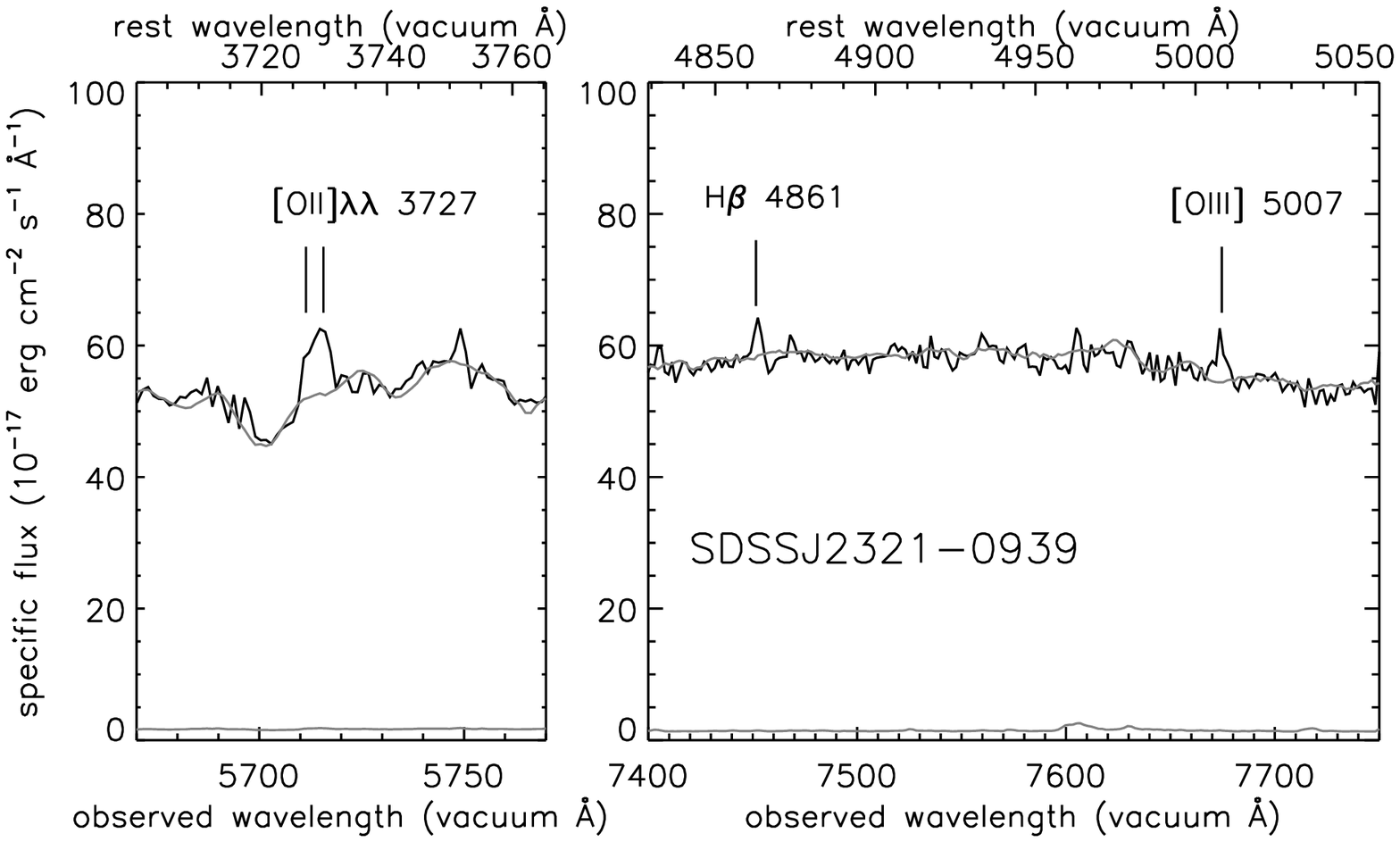}{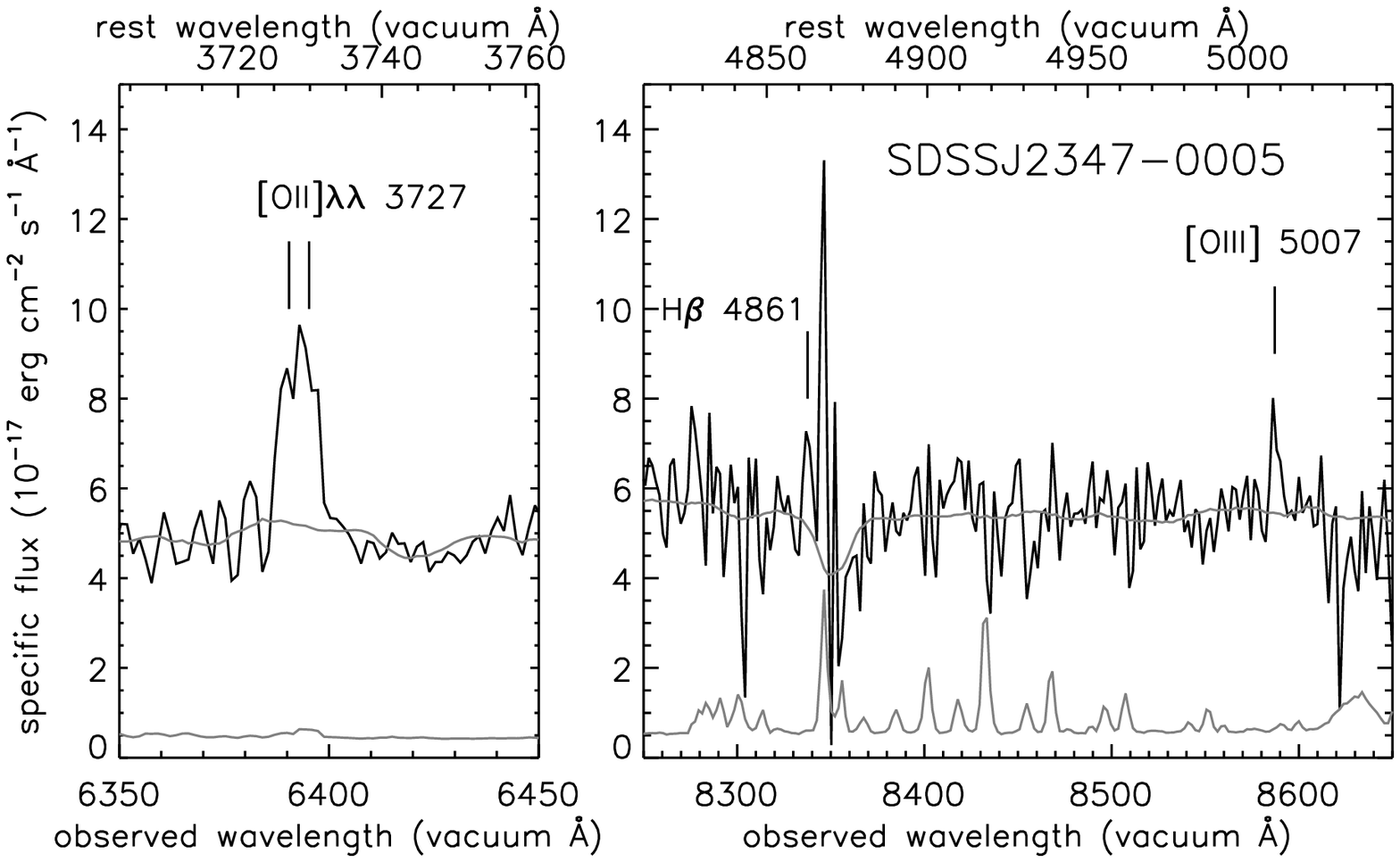}}
\caption{(continued)}
\end{figure*}

\subsection{Candidate Selection}
\label{sec:selection}

The method by which we select our lens candidates is essentially that
described by B04.  Briefly, we
subtract best-fit PCA templates (a byproduct of the redshift pipeline)
from the observed SDSS target-galaxy spectra.  We require the SDSS
continuum to be well-fit by the template, which effectively yields a
parent sample of galaxies with well-behaved absorption-dominated
spectra and very secure redshifts, which we denote $z_{\mathrm{FG}}$
(for ``foreground'').  The residual spectra are then scanned for
nebular line emission at redshifts greater than $z_{\mathrm{FG}}$.
Spectra in which such emission is significantly detected for at least
three separate common atomic transitions at a single background
redshift ($z_{\mathrm{BG}}$) are taken as lens candidates for having
emission at two different redshifts along the same line of sight as
sampled by the 3-arcsec diameter SDSS spectroscopic fiber.

Taking $z_{\mathrm{FG}}$, $z_{\mathrm{BG}}$, and the measured stellar
velocity dispersion $\sigma_a$ from SDSS allows us to determine to
first order the strong-lensing cross section for each system in the
source plane -- using a singular isothermal sphere (SIS) model -- as
$\pi \theta_E^2$, with the Einstein radius given by $\theta_E = 4 \pi
(\sigma_a^2 / c^2) (D_{LS} / D_S)$. In the image plane the region of
multiple imaging is four times larger than the source-plane
strong-lensing cross section, in the sense that
all images within a radius $2 \theta_E$ of the potential center
will show strongly-lensed counterimages. To maximize the number of
strong lenses in our survey, we formed our Cycle--13 {\sl HST} target
list from the candidates with the highest predicted lensing cross
section.  Thus our {\sl HST} target sample is velocity-dispersion
selected to leading order, with an additional selection bias in favor
of systems with larger angular-diameter distance ratios $D_{LS} / D_S$
(a function of the redshifts $z_{\mathrm{BG}}$ and $z_{\mathrm{FG}}$).
We take 20 of our targets from the list of lens candidates published
in B04, which were selected from within the luminous red galaxy
spectroscopic sample of the SDSS \citep[LRG,][]{eisenstein_lrg}.  The
LRG sample is defined by photometric selection cuts that very
efficiently select massive early-type galaxies in the redshift range
$0.15 < z_{\mathrm{FG}} \la 0.5$, as confirmed by SDSS spectroscopy.
These galaxies are very homogeneous in their spectral, photometric,
and morphological properties, and we place no further requirements on
our LRG-sample lens candidates beyond their photometric selection as
described in \citet{eisenstein_lrg} and subsequent spectroscopic
confirmation as galaxies with redshifts $z > 0.15$.  The remaining 29
targets on our Snapshot list are selected with the same spectroscopic
algorithm from within the MAIN galaxy sample of the SDSS
\citep{strauss_main}.  The MAIN sample is much more heterogeneous, and
we impose a quiescent, absorption-dominated spectral criterion by
requiring our lens candidates to have rest-frame equivalent widths in
H$\alpha$ of $EW_{\mathrm{H}\alpha} < $1.5\,\AA\ (with a few
exceptions made in the interest of maximizing the total number
discovered lenses in the program: see SDSSJ1251-0208 below.).
Although most of the systems we target do indeed exhibit early-type
morphology, the selection of SLACS lens candidates from within the LRG
and MAIN galaxy samples of the SDSS is purely spectroscopic.

\subsection{ACS Image Processing}
Our ACS Snapshot observing strategy is discussed in
\citet{bolton_1402}, and consists of one 420s Wide-Field Channel (WFC)
exposure through each of the two filters F435W and F814W\@.  The SLACS
lens candidates are selected to yield bright lenses with faint
background sources.  This facilitates the photometric, morphological,
and kinematic study of the lens galaxy, but can also make the
relatively faint lensed features difficult to detect and even more
difficult to use in lens modeling.  The key to success is effective
subtraction of the image of the lens galaxy by fitting a smooth model
to the image data of the candidate lens galaxy
\citep[e.g.][]{peng_galfit, simard_gim2d}.  Given the extremely
regular isophotal structure of most early-type galaxies, this is a
reasonable proposition.  The most common parameterized model for
early-type brightness distributions is the generalized de Vaucouleurs
or S\'{e}rsic law \citep{sersic_law,ciotti_sersic}.  We first
attempted to use elliptical S\'{e}rsic model-fitting and subtraction
in our analysis, but generally found that the global systematic
residuals of the fit were large compared to the surface-brightness
levels of the lensed features that we hope to use to constrain
gravitational-lens models.  Significant galaxy-core residuals are
especially pronounced in the S\'{e}rsic-subtracted F814W data.  The
shortcomings of the S\'{e}rsic model led us to implement a more
generalized galaxy-model fitting procedure involving a b-spline fit to
the radial profile with a low-order multipole dependence to fit the
angular structure, which we describe in Appendix~\ref{bsplineapp}.

\section{NEW LENSES AND OTHER OBSERVED SYSTEMS}
\label{lenscat}

Here we present the catalog of 28 candidate lens systems\footnote{The
Cycle--13 program is still ongoing and another 11 systems have been
observed, with 9 more being scheduled.} observed by the SLACS survey
through 2005 March 22.  Photometric and spectroscopic parameters for
the sample as measured by SDSS are given in Table~\ref{galtab}.
Figure~\ref{specfig} presents the SDSS discovery spectra of all
targeted systems, focused on the background-redshift line emission.
For definiteness, we enumerate the several possible explanations
for any one of our spectroscopically selected lens candidates as follows:
(1) a multiply imaged background galaxy (``strong lens''),
(2) a singly imaged though possibly magnified background galaxy
(a ``non-lens'' in our classification), (3) a multiple-image
system corresponding to a projection of multiple singly imaged
sources with similar colors, and
(4) spurious noise features in the SDSS spectrum.
Given the adopted significance threshold for selection and
the careful treatment of the noise properties of SDSS
spectra (B04), we expect (4) to be an unlikely explanation,
though it merits consideration in the case of any lower
signal-to-noise candidates that do not produce detections
in follow-up observations.  Systems falling into category
(3) could in principle be mistaken for bona fide strong
lenses, but can be expected to have ill-fitting lens
models, since real lenses occupy only a small subspace
of all conceivable image configurations.
For simplicity, this paper only categorizes ACS-observed
systems into ``strong lenses'' on one hand and ``everything else''
on the other, with ``everything else'' including systems
for which a definitive explanation cannot be made
based on the data in hand (i.e.\ ACS non-detections).

To determine the incidence of strong lensing, we examine the F814W and
F435W residual images formed by subtracting smooth models of the
foreground galaxy constructed as described in
Appendix~\ref{bsplineapp}.  If the residual image of a target shows
multiple images with similar colors that can be reproduced by a simple
lens model with the potential center fixed at the optical center of
the foreground galaxy we classify the system as a strong gravitational
lens.  The lens models for these systems are presented in Paper III,
but we summarize the lens-modeling process here.
The procedure is based on the method of \citet{warren_dye_03},
with an implementation as described in \citet{tk04} and
\citet{koopmans_2005}.
We parametrize the lens galaxy as a singular isothermal ellipsoid
\citep[SIE,][]{kormann_sie}, with three free parameters
corresponding to lens strength (i.e.\ Einstein radius),
ellipticity, and major-axis position angle.  The source-plane
brightness distribution of the candidate lensed galaxy is
represented on a pixel grid.  We determine the best-fit lens
parameters and source-plane pixel brightnesses
by minimizing $\chi^2$ with respect to the galaxy-subtracted
$I$-band
image-plane data, while also imposing a brightness-dependent
regularization on the source-plane distribution to
suppress curvature.
Unsuccessful strong-lens models are characterized by an inability
to map putative multiply-imaged features to a single
region of the source plane for any set of SIE model parameters.
SDSSJ1025 is an example of a system with an apparent
arc and counterimage which nonetheless cannot be reproduced
by an SIE lens model, and is thus not classified as a
strong lens.
ACS data for target galaxies classified as lenses are shown in
Figure~\ref{acsfig}.  We note again that {\em all} target systems have
line emission in their SDSS spectra at a redshift higher than that of
the foreground galaxy, and thus the lensing interpretation does not
rest on imaging data alone.

Figure~\ref{maybefig} presents all systems other than those that we
classify solidly as lenses.  This figure encompasses systems that may
indeed be strong lenses as well as systems that appear to be definite
{\em non}lenses.  We note that a significant number of systems that we
consider possible but not definite lenses show a faint candidate
counterimage near the center of the foreground galaxy opposite a more
prominent image at larger radius.  If these features were all due to
simple residual error in the foreground galaxy subtraction, this
configuration would not be expected.  The features could conceivably
be explained by the prominent image ``pulling'' the foreground-galaxy
model to the side, leading to an under-subtraction of the
foreground-galaxy flux near the center on the opposite side.  We
believe this explanation is unlikely since we have taken care to mask
neighboring images when fitting the foreground-galaxy model
(Appendix~\ref{bsplineapp}), and we do not see any corresponding
over-subtraction on the near side of the core.  The most asymmetric
double lenses will in general have faint counterimages at small
angular offset from the lens center and will be the most difficult
lenses to confirm, although their status as lenses or non-lenses can
have a significant impact on statistical inferences based on the lens
sample.  Integral-field spectroscopy provides the best chance to
associate background-redshift line emission with these faint putative
counterimages near the lens center
and thereby confirm or reject a strong-lensing hypothesis.

Many of the lens galaxies shown in Figure~\ref{acsfig} show such
striking features in their residual images that their interpretation
as strong lenses is effectively certain when considered in combination
with their SDSS spectroscopic detection and gravitational-lens models
(Paper III).  For putative lenses with less dramatic morphology,
further evidence as to their status may be obtained with
integral-field spectroscopy, which can confirm or deny the spatial
coincidence between the high-redshift line emission and the candidate
lensed features seen with {\sl HST}.  For several of our SLACS target
galaxies, we have obtained spatially resolved spectroscopy with the
integral-field units (IFUs) of the Inamori Magellan Areal Camera and
Spectrograph \citep[IMACS,][]{bigelow_imacs, schmoll_imacs_ifu} on the
6.5m Walter Baade (Magellan I) telescope at Las Campanas Observatory
and the 8m Gemini-North Multi-Object Spectrograph
\citep[GMOS-N,][]{hook_gmos, murray_gmos_ifu} at Mauna Kea; we present
these data in Figure~\ref{slacs_ifu}.  In all cases, the IFU
emission-line images are coincident with the ACS residual-image
morphology, supporting the strong-lensing interpretation.

We note here that the background sources in our lens
systems are generally faint galaxies with irregular
morphology.  As evidenced by the emission-line flux
by which they are originally detected, they should also be
regarded as starforming.  The absence of significant broadening
of the Balmer emission lines at the $\sim 150$km\,s$^{-1}$
resolution of the SDSS further suggests that they do
not host quasars.  For a recent observational study of the
likely source population of emission-line galaxies,
see \citet{droz_grism}.

\begin{figure*}[t]
\plotone{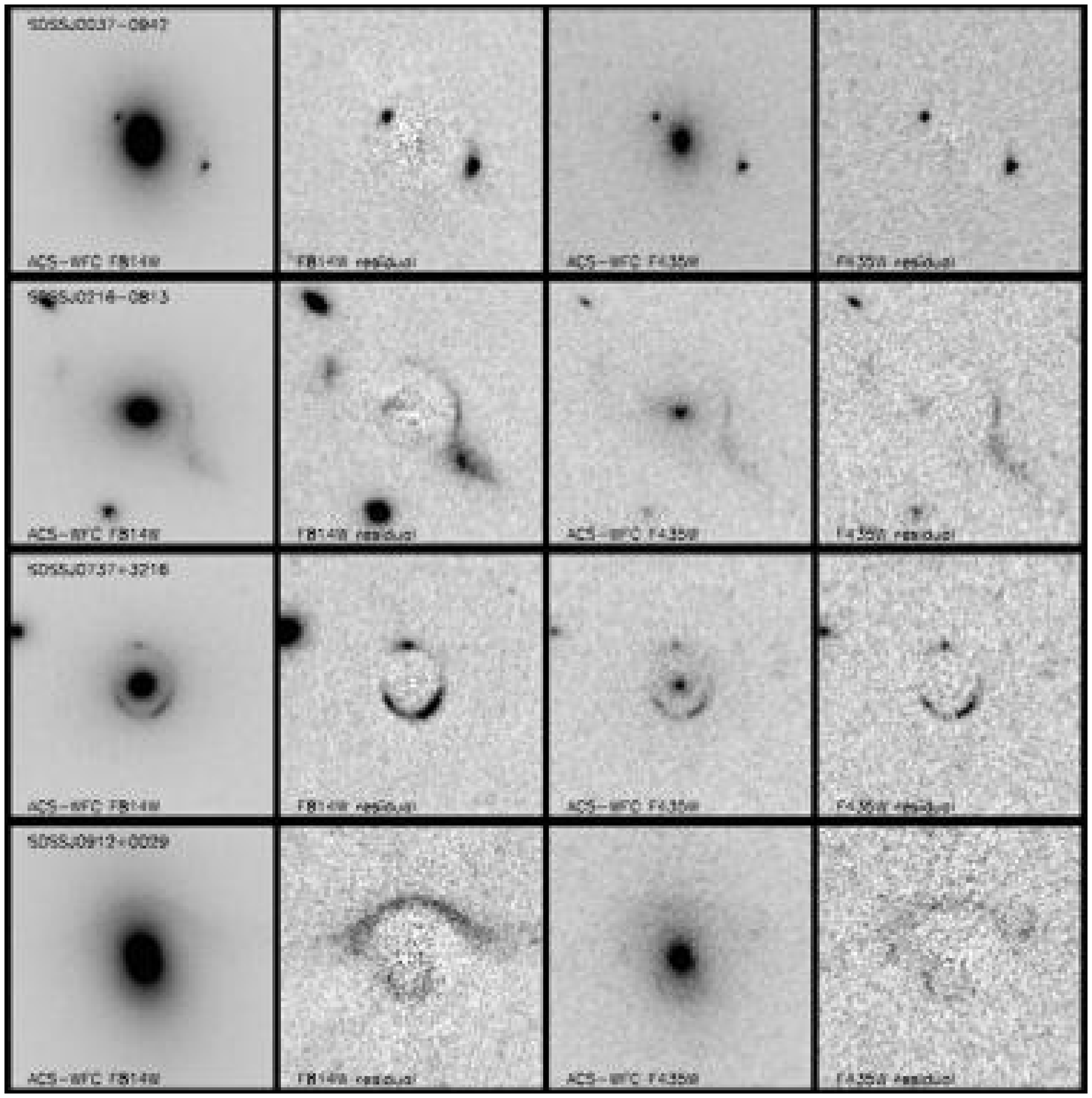}
\caption{\label{acsfig} {\sl HST} ACS-WFC F814W and F435W imaging of
new gravitational lenses from the SLACS survey.  Also shown are
residual images with smooth b-spline lens-galaxy models subtracted,
revealing lensed features more clearly.  Images are formed from
flat-fielded single-image native ACS data.  Cosmic-ray and other
zero-weight pixels are replaced with values from a median-smoothed
residual image, with the b-spline model galaxy values added for the
direct images.  Images are $8\arcsec \times 8\arcsec$, with N up and E
left.  All systems in this figure are modeled successfully with
singular isothermal ellipsoid lens models, as described in Paper III.
}
\end{figure*}

\addtocounter{figure}{-1}

\begin{figure*}[t]
\plotone{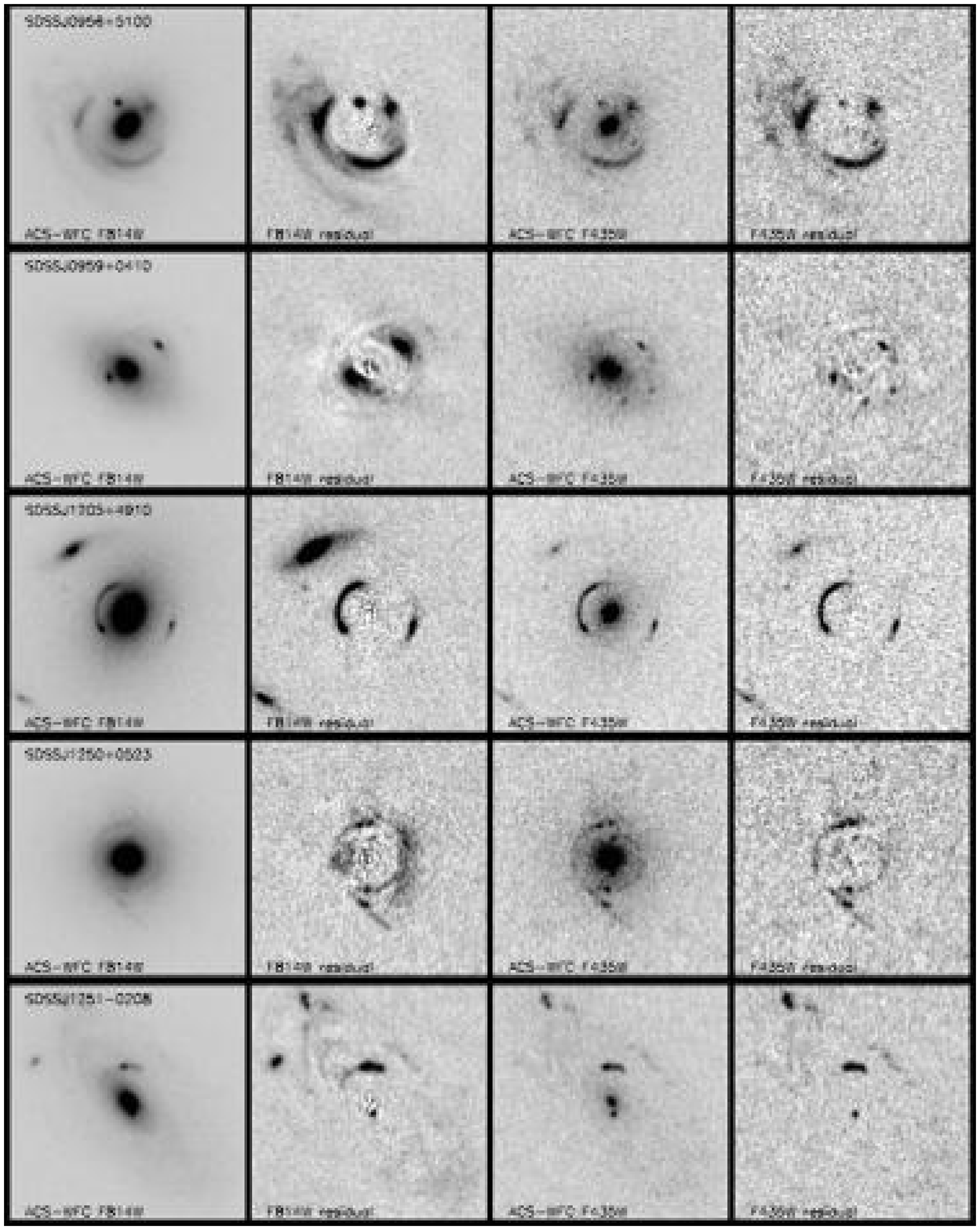}
\caption{(continued)}
\end{figure*}

\addtocounter{figure}{-1}

\begin{figure*}[t]
\plotone{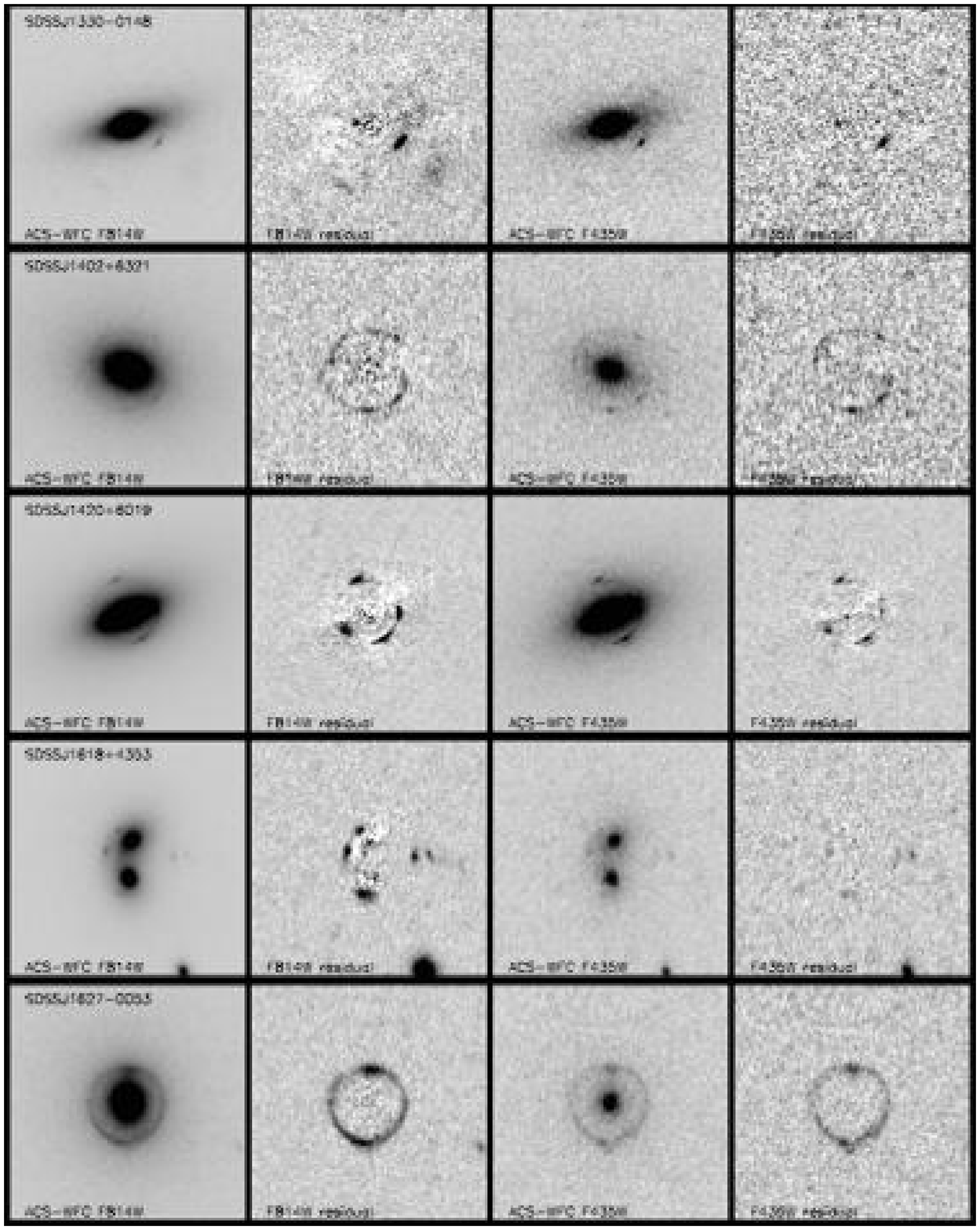}
\caption{(continued)}
\end{figure*}

\addtocounter{figure}{-1}

\begin{figure*}[t]
\plotone{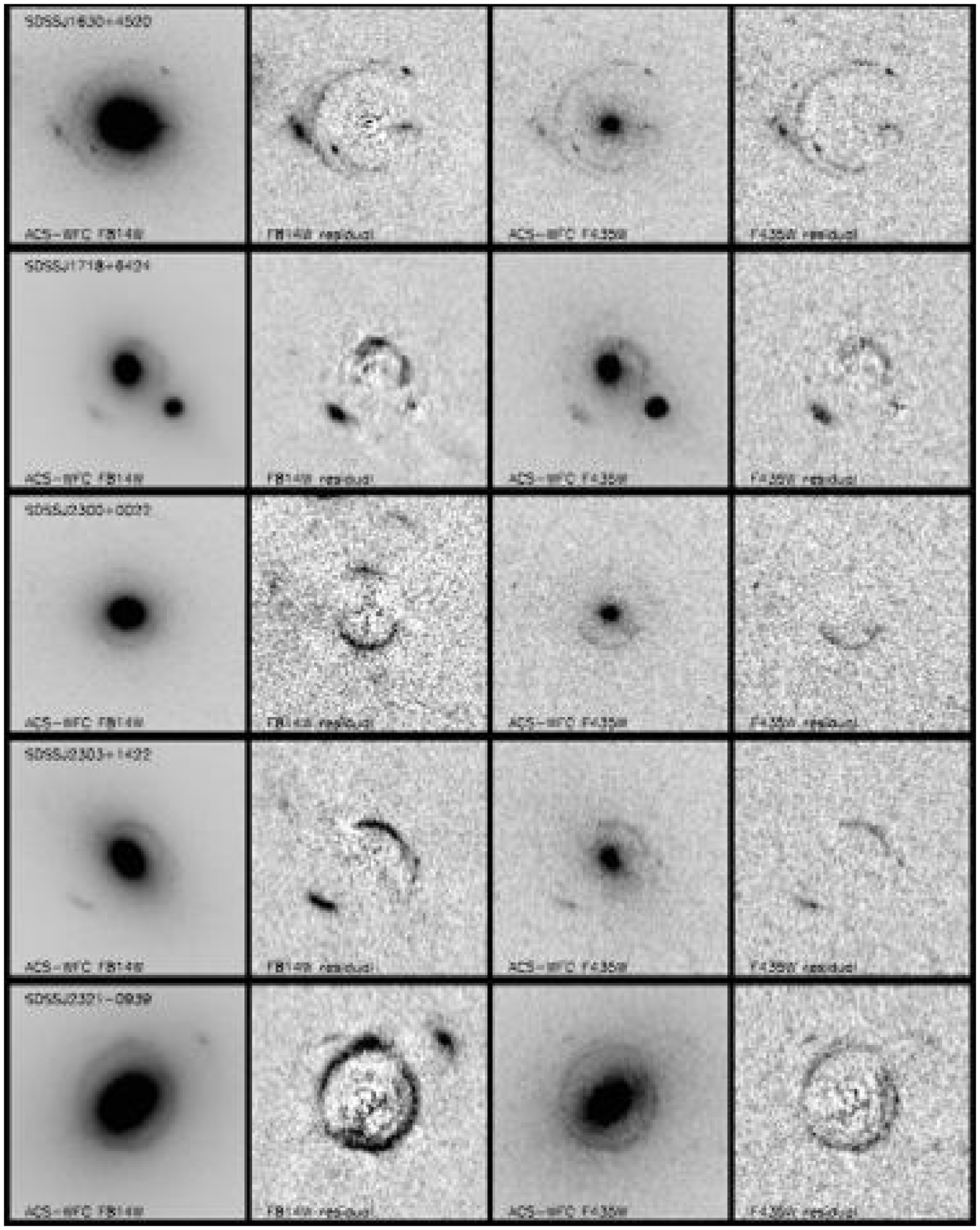}
\caption{(continued)}
\end{figure*}

\begin{figure*}[t]
\plotone{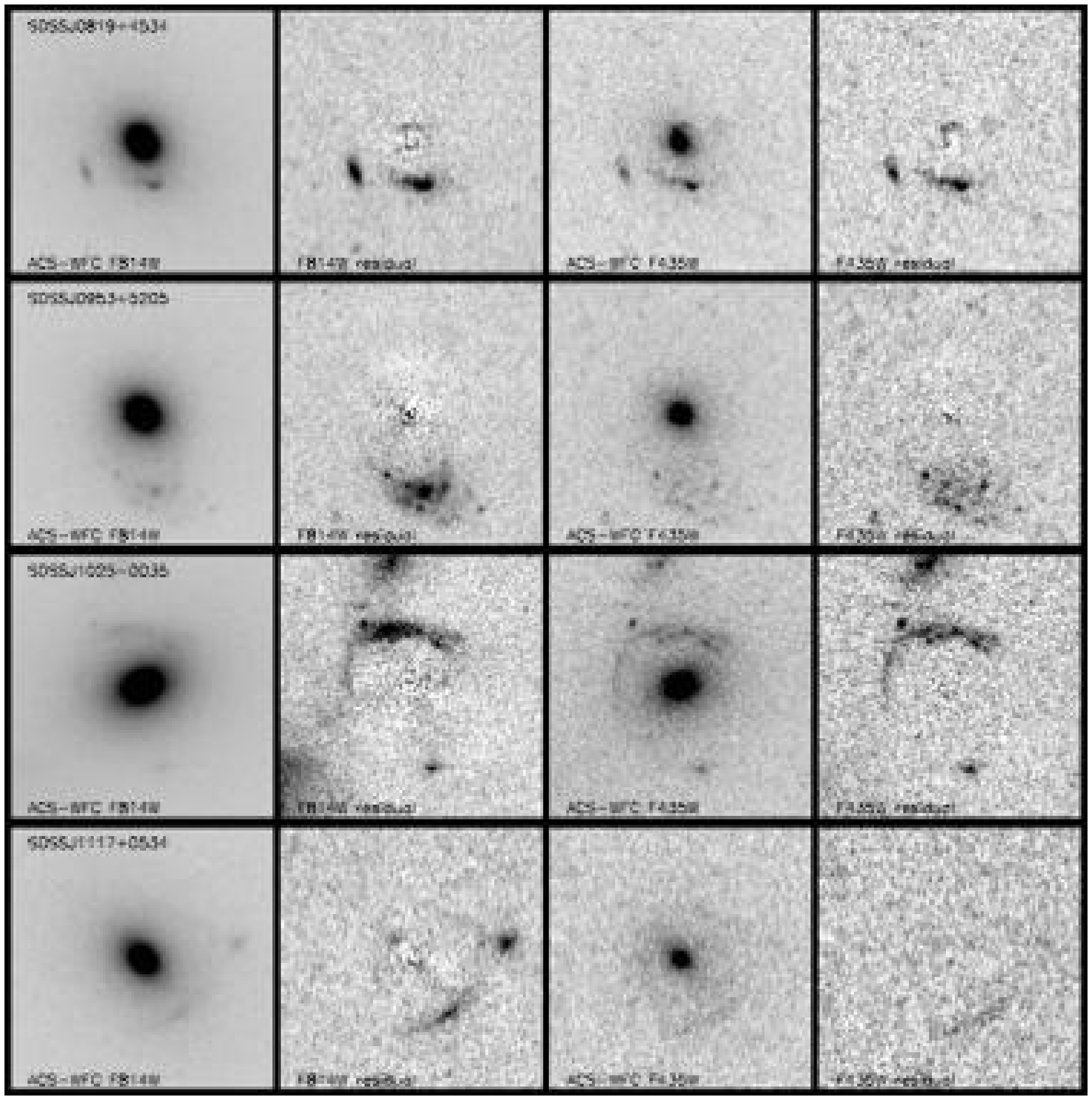}
\caption{\label{maybefig} {\sl HST} ACS-WFC imaging of lens candidates
observed by the SLACS survey for which the ACS data are inconclusive
as to the incidence of strong lensing.  Images are as in
Figure~\ref{acsfig}.  See notes on individual systems in
Appendix~\ref{maybeapp}}
\end{figure*}

\addtocounter{figure}{-1}

\begin{figure*}[t]
\plotone{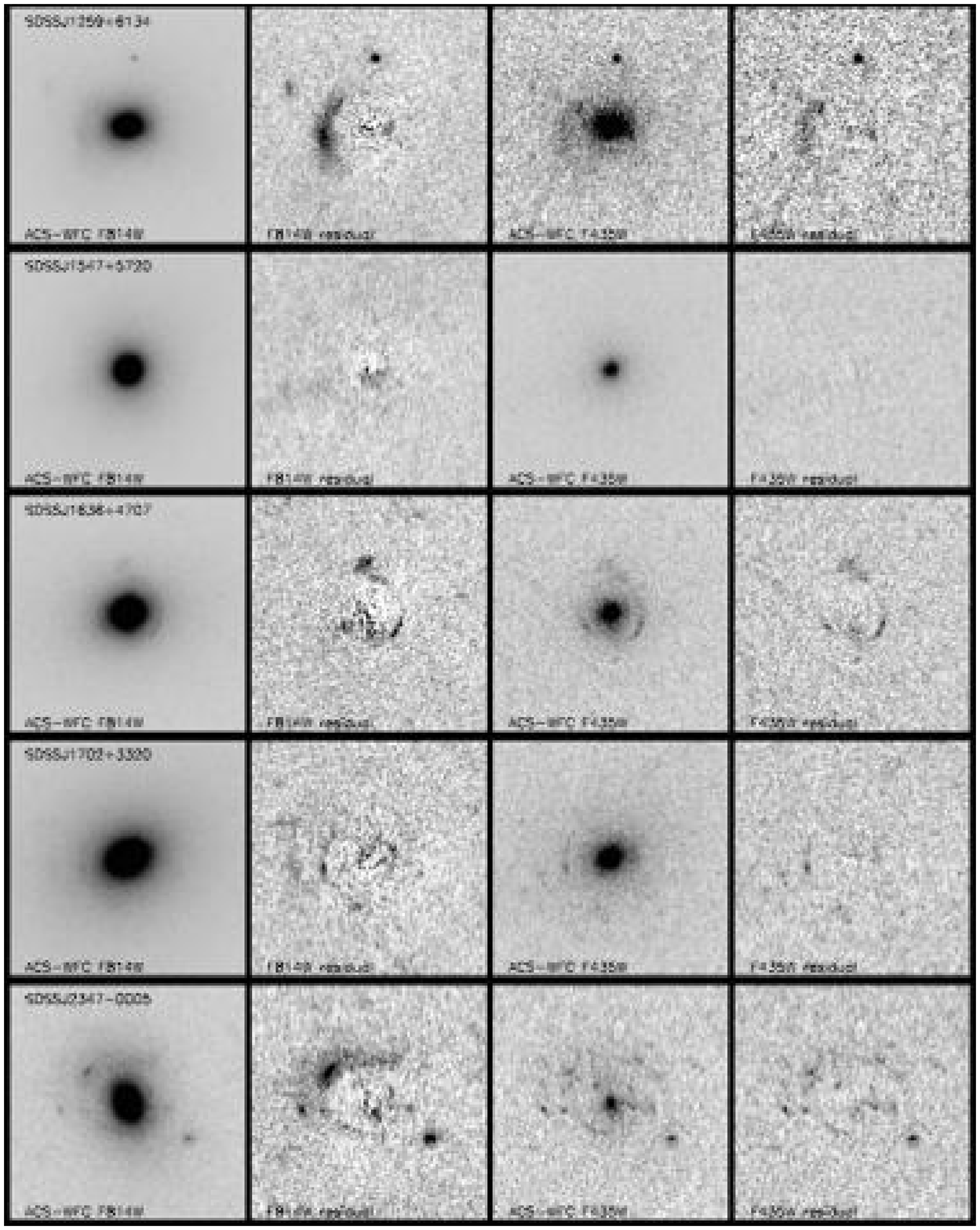}
\caption{(continued)}
\end{figure*}

\section{STATISTICS AND SELECTION EFFECTS}\label{sec:stats}

The statistics of strong gravitational lensing are the subject of a
great body of literature \citep[e.g.][]{turner_lensstat, Fukugita_1991,
kochanek_lensstat_96, chae_lensstat_02}, most of it dealing with the
lensing of quasars by galaxies.  The rigorous statistics of the SLACS
survey, though involving similar considerations, would be sufficiently
distinct to require their own detailed treatment: our original sample
is formed from potential lensing objects rather than from potentially
lensed sources, and we must consider the effects of an extended source
image and finite fiber sampling.  We defer a full lens-statistical
analysis of the SLACS survey to a future paper. Here we address the
selection effects of the SLACS survey in terms of the robust
statistics of the distribution of lens-galaxy observables.

As discussed above, the SLACS target selection involves an explicit
velocity-dispersion and redshift selection.  This should not in itself
present any complication in the comparison of our lenses to other
galaxies with similar velocity dispersions and redshifts.  After
controlling for this selection, we may ask whether our lenses are a
representative sample of the populations from which they have been
drawn: luminous red galaxies and quiescent main-sample galaxies from
the SDSS.  If so, their relation to early-type galaxies selected by
other means is simply determined by the SDSS LRG and MAIN sample
target selection function \citep{eisenstein_lrg, strauss_main}.  The
question can be framed between two alternatives: do we select simply
for spectral superposition caused by
a chance foreground-background coincidence,
or is there some significant bias by which galaxies with certain
properties are more likely to be selected as lens candidates based on
their SDSS spectra?  To test this possibility we exploit the parent
sample of $\sim10^5$ SDSS galaxies (Early Data Release through Data
Release 3).  For each lens we construct a control sample of SDSS
galaxies with the same redshift and velocity dispersion (within the
uncertainties)---quantities for which we explicitly select---and test
whether the lens galaxy is typical of galaxies in the control sample
in terms of its magnitude, color, effective radius, and isophotal
ellipticity---quantities for which we do not explicitly select.
Although this is not a definitive test of the representative nature of
our lenses (it will not uncover any ``hidden variable'' bias), it is
straightforward, robust, and informative.  By forming control samples
at the redshift of the lenses, we avoid the necessity of applying
evolution- and $k$-corrections to the broadband magnitudes.  We may
perform our comparison sensibly for galaxies with velocity dispersions
well-measured by SDSS.  We exclude the lens SDSSJ1251$-$0208 from the
analysis, both because its spectral signal-to-noise ratio is too low
for a confident velocity-dispersion measurement and because its
H$\alpha$ equivalent width exceeds our quiescent threshold.  (In fact,
it is a bulge-dominated spiral-galaxy lens).  The double-lens galaxies
SDSSJ1618$+$4353 and SDSSJ1718$+$6424 are also excluded, as is the
lens SDSSJ1205$+$4910, which has significant flux from a neighboring
galaxy within the SDSS spectroscopic fiber.  This leaves a
well-defined sample of 15 single early-type lenses.  The comparison
sample for each lens consists of unique galaxies from the SDSS
database with redshifts within $\pm 0.005$ of the lens redshift,
velocity dispersions within $\pm 15$\,km\,s$^{-1}$ (approximately the
median velocity-dispersion error of the sample) of the lens galaxy,
and median signal-to-noise per spectral pixel greater than 8.  For
LRG-sample targets, we also require the comparison sample to pass the
photometric LRG cuts, and for the MAIN-sample lenses, we require a
rest-frame $EW_{\mathrm{H}\alpha} < 1.5$ \AA\@.  We make a
luminosity-distance correction to the broadband magnitudes and an
angular-diameter-distance correction to the effective radii within the
redshift slice to place all galaxies closer to the exact redshift of
the lens.  The resulting individual comparison samples have as few as
17 and as many as 1793 galaxies, with a mean of 451 and a median of
337 control galaxies per lens galaxy.

We would like to answer the question of whether our lenses have
observables (luminosities, colors, effective radii, and ellipticities)
consistent with having been drawn from the distribution of those
observables seen in the control samples.  Since we only have one lens
at each redshift, a straightforward Kolmogorov-Smirnov \mbox{(K-S)}
test of the lens sample against the control samples is not possible:
even under the null hypothesis, each individual lens has a different
parent distribution.  We may however put all these parent
distributions on common footing by using the fact that the \mbox{K-S}
test is invariant under a monotonic rescaling of the variable under
consideration.  Specifically, for each individual lens's control
sample, there exists a transformation of the observable of interest
that converts the control distribution into a uniform distribution
between the minimum and maximum values.  The correspondingly
transformed lens observable is simply equal to the normalized rank
(between 0 and 1) of the lens quantity within the cumulative
distribution of the control sample in that quantity.  Thus we may
perform a one-dimensional \mbox{K-S} test of the distribution of
lens-observable ranks against a uniform distribution over the interval
0 to 1.  This test is in some sense like a rank-correlation test
within the \mbox{K-S} formalism: does ranking as a lens correlate with
ranking in luminosity, color, size, or ellipticity?

Figure~\ref{ksplot} shows the cumulative distributions of lens
rankings in $r$ (magnitude), $g-r$ (color), $R_e$ (effective radius),
$b/a$ (isophotal axis ratio), and $I_e$ ($r$-band effective surface
brightness: $I_e \propto 10^{-0.4 r} R_e^{-2}$) within their control
samples.  From these distributions we compute the statistic
$D_{\mathrm{KS}}$, equal to the maximum difference between the
cumulative rank distribution and the null-hypothesis distribution.
Since the parent distribution under the null hypothesis is known
exactly by construction (uniform probability for any rank between 0
and 1), and since the total number of control galaxies is much larger
than the number of lenses, we compute the significance of
$D_{\mathrm{KS}}$ for a distribution of 15 values against a known
parent distribution \citep[e.g.][]{press_nr}.  (The statistical
significance of an outlying lens will be limited by a smaller control
sample in that its rank is at least $1/N_{\mathrm{gal}}$ and at most
$1 - 1/N_{\mathrm{gal}}$, where $N_{\mathrm{gal}}$ is the number of
galaxies in its control sample.)  The resulting probabilities of the
lens sample having been drawn at random from the control-sample
distributions in $r$, $g-r$, $R_e$, $b/a$, and $I_e$ are 0.097, 0.655,
0.264, 0.550, and 0.085 respectively.  In terms of color and axis
ratio, the SLACS lenses seem to be a representative sample of the
parent distribution of SDSS galaxies.  However, the distribution of
SLACS galaxies in brightness and effective radius (and hence in
surface brightness) is somewhat significantly skewed towards brighter
and more centrally concentrated systems.

\begin{figure*}[t]
\centerline{\scalebox{0.7}{\plotone{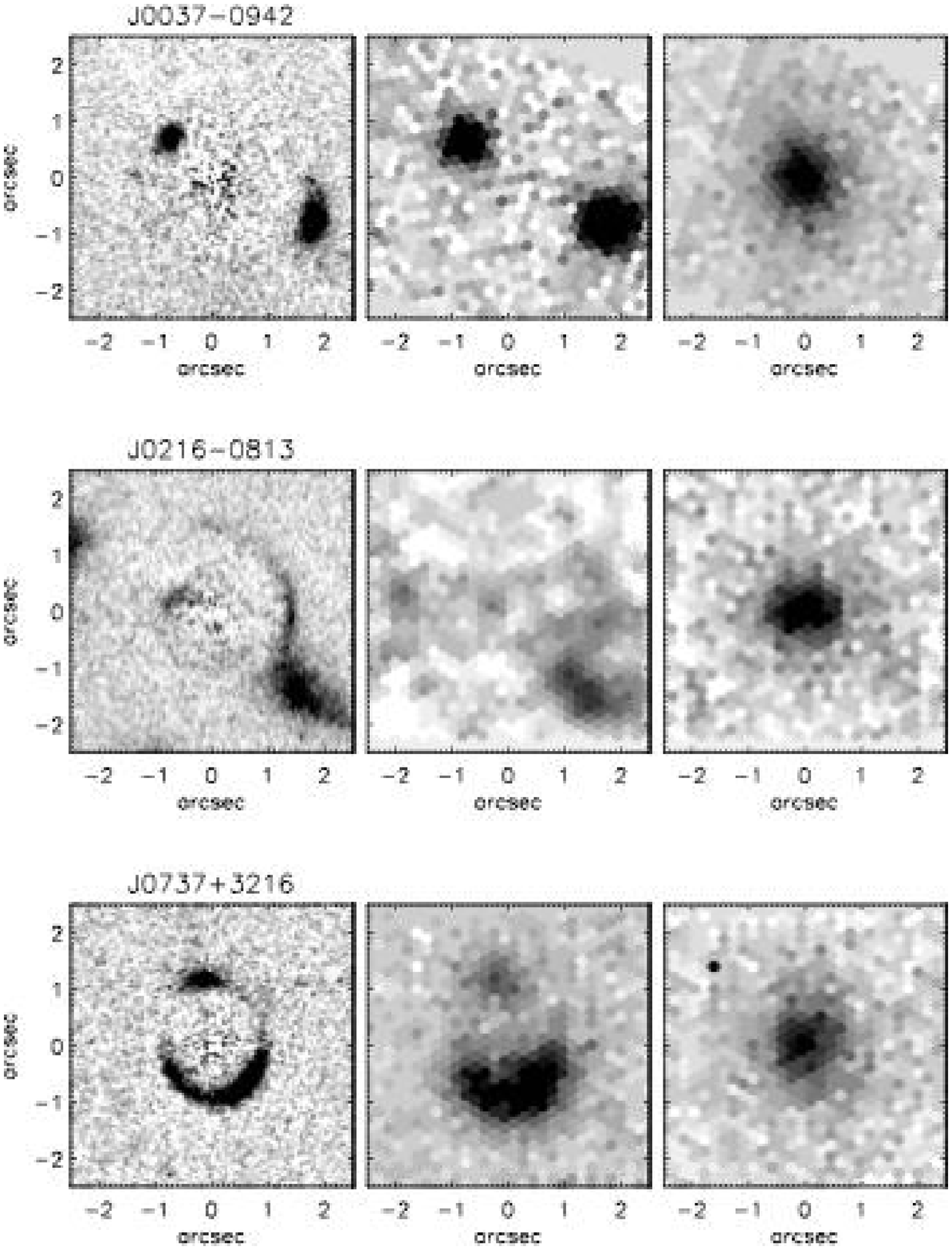}}}
\caption{\label{slacs_ifu} IFU narrowband imaging of several SLACS
target systems.  Left panels show {\sl HST}-ACS F814W residual images,
for reference.  Center panels show IFU emission-line images, and right
panels show IFU continuum images.  IFU images are constructed from IFU
spectra by fitting linear continuum and Gaussian emission-line models
as described in Bolton \& Burles (in preparation).  Data are from
IMACS (SDSSJ0037, SDSSJ0216, and SDSSJ2321) and GMOS-N (others).
Emission-line images for SDSSJ0216, SDSSJ0956, and SDSSJ1547, have
been smoothed spatially with a 7-lenslet hexagonal kernel to suppress
noise.  Narrowband images are formed at the redshifted wavelength of
the following background-galaxy emission lines: \oiiib\ (0037, 0737,
1402, 1702), \hb\ (0956), and \oii\ (0216, 1547, 1630, 2321)}
\end{figure*}

\addtocounter{figure}{-1}

\begin{figure*}[t]
\centerline{\scalebox{0.7}{\plotone{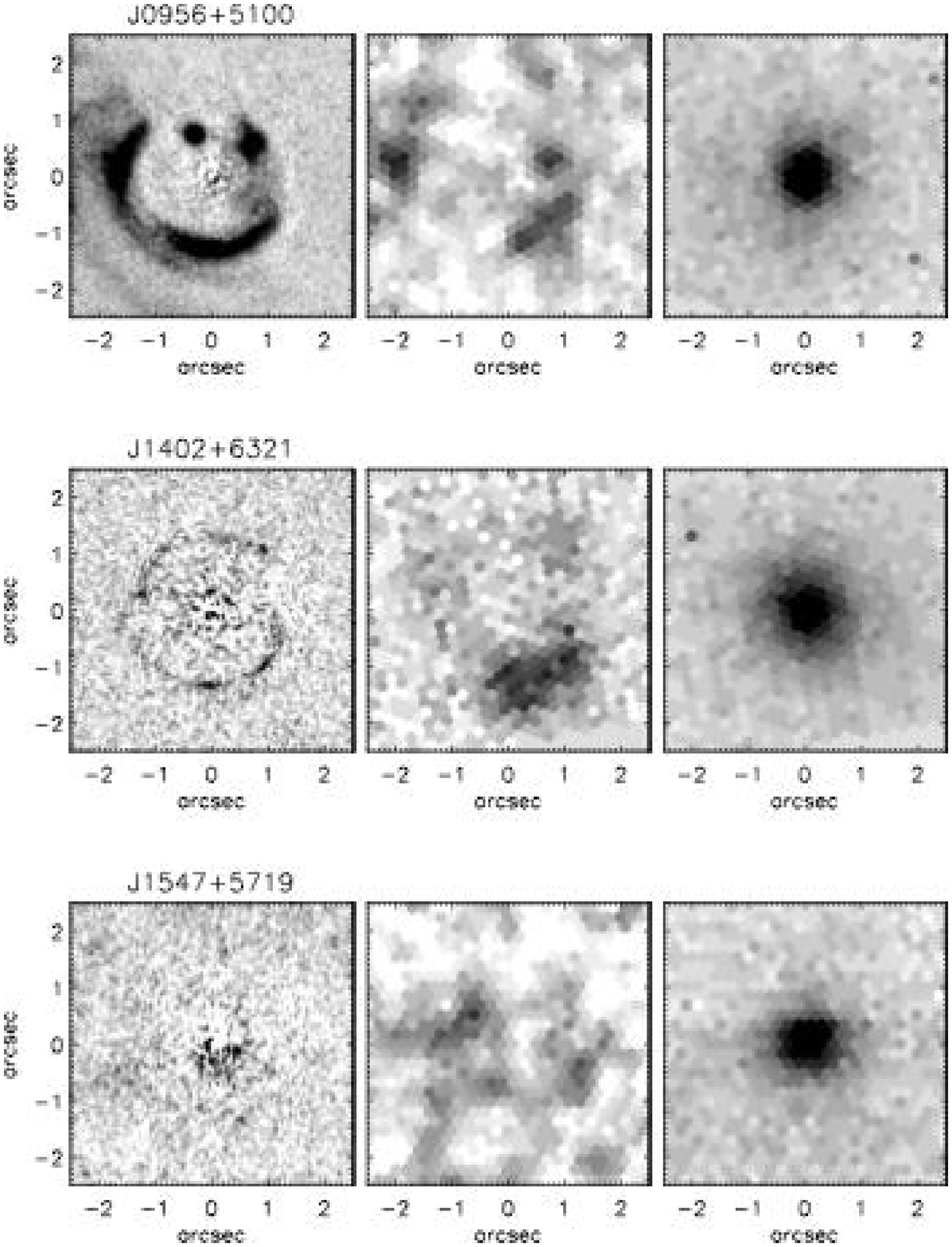}}}
\caption{(continued)}
\end{figure*}

\addtocounter{figure}{-1}
\begin{figure*}[t]
\centerline{\scalebox{0.7}{\plotone{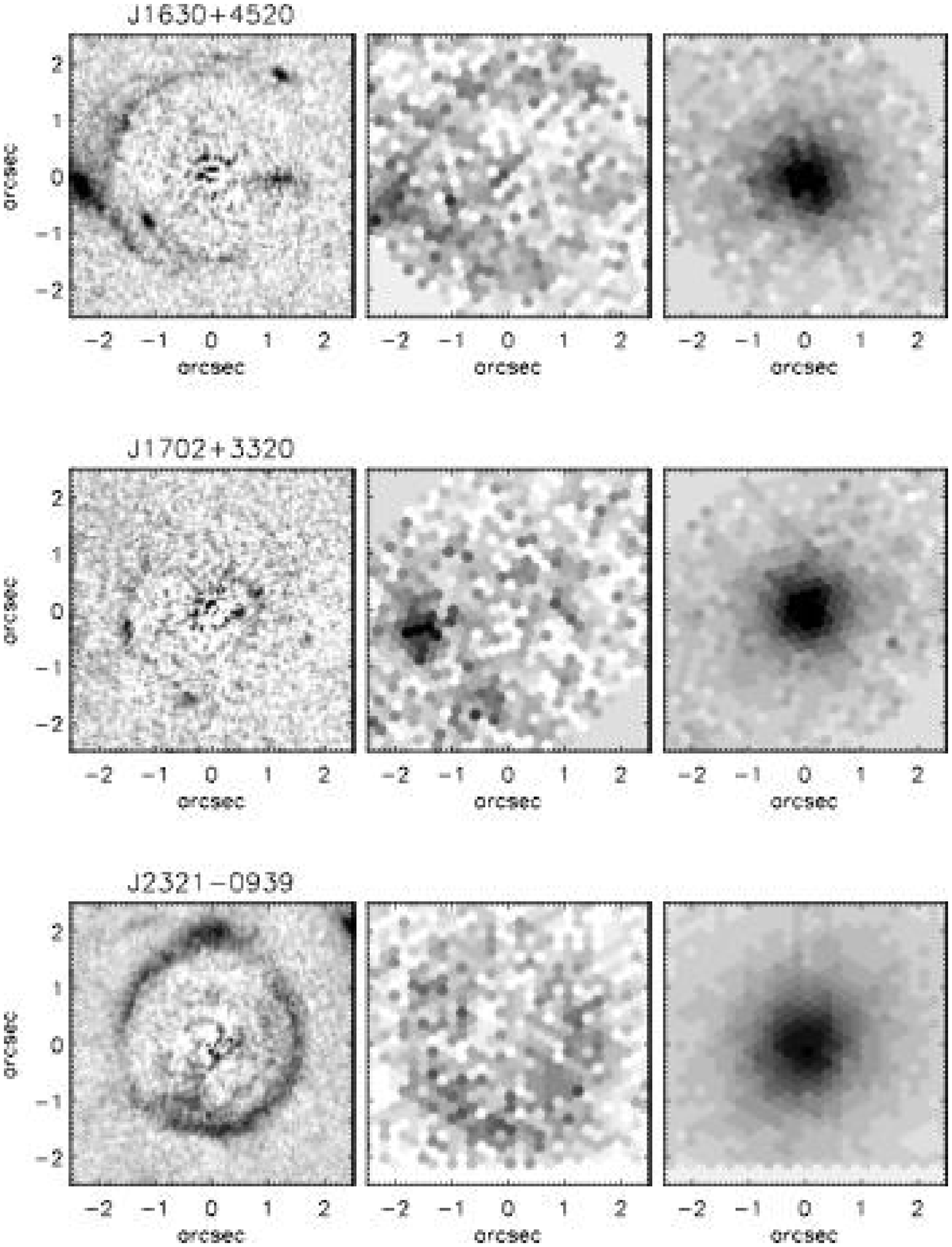}}}
\caption{(continued)}
\end{figure*}

This trend is in the same sense as that
of the FP: at fixed velocity dispersion, brighter galaxies have
smaller effective radii. Paper II shows that the SLACS lenses fall on
the local FP corrected for stellar evolution, and further examines the
degree of lens bias towards higher effective surface brightness in the
context of the FP.  Although the statistical significance is not
great, the trend is at first glance suggestive of the effect that
more centrally condensed objects are more efficient
gravitational lenses \citep[e.g.][]{li_ostriker},
to the extent that stellar mass constitutes the dominant
lensing component.  A larger sample of similarly selected lenses will
allow us to test this selection effect with greater statistical
significance.  Possible interpretations of this finding---as discussed
in detail in Papers II and III---include observational selection
effects due to the finite size of the SDSS fiber which may bias our
survey towards the highest surface brightness lens galaxies.
In Paper III we will see that, within the context of powerlaw
mass models, the SLACS lens sample shows great
homogeneity in the slope of the total mass density profile
($\gamma'$) and no significant
correlation between $\gamma'$ and $I_{\rm e}$.  This
uniform degree of mass concetration within the
lens sample itself argues somewhat against a lens-selection
bias due to mass concentration, but this argument
is limited by the lack of an equally direct
probe of the mass distribution in
the non-lens control sample.

\begin{figure*}[t]
\centerline{\plotone{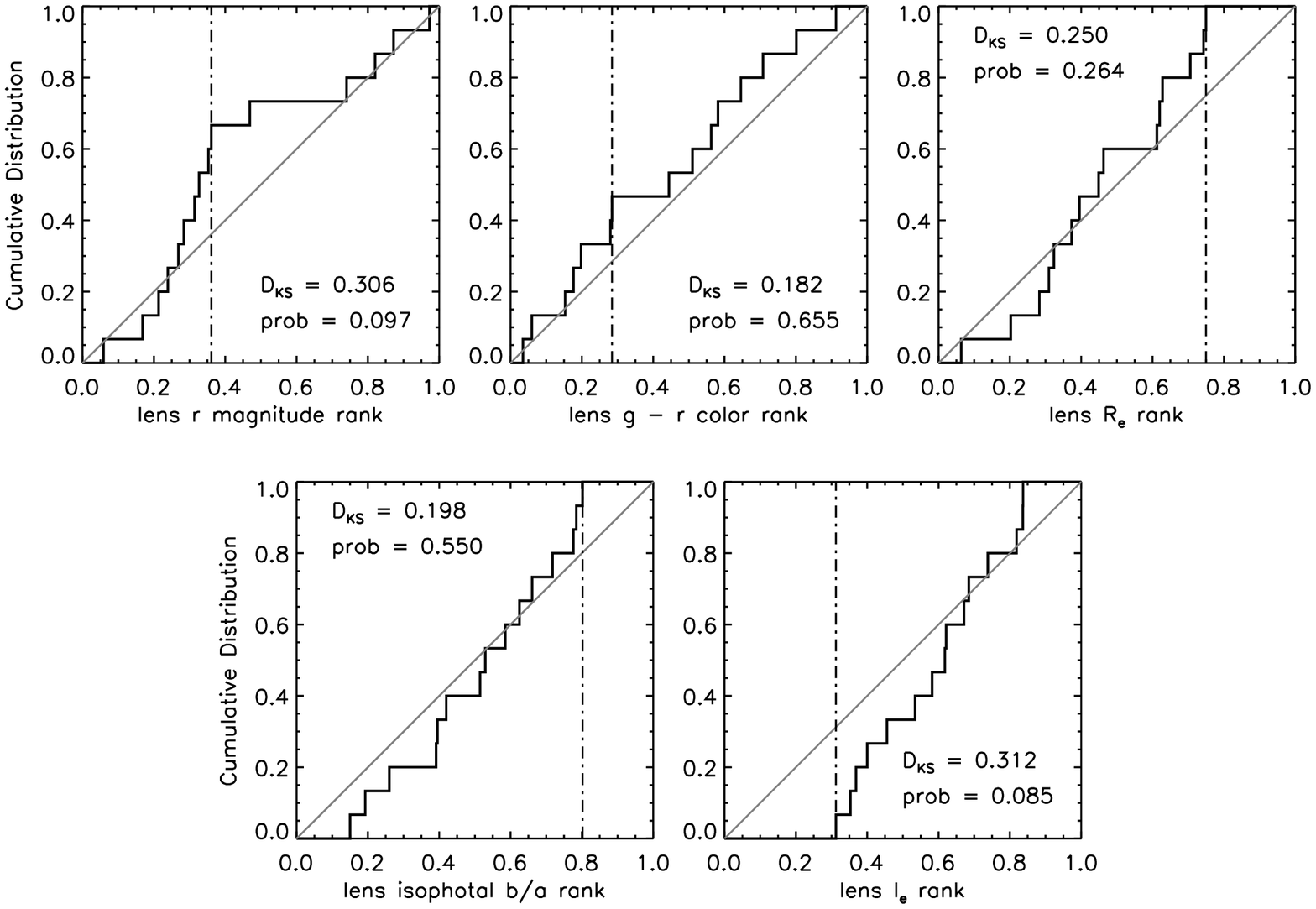}}
\caption{\label{ksplot} Kolmogorov-Smirnov tests of the rank of
lens-galaxy observables within the cumulative distributions of those
observables in control galaxy samples with similar redshifts and
velocity dispersions.  From upper left to lower right are the
distributions for $r$-band magnitude, $g-r$ color, effective
(half-light) radius $R_e$, isophotal axis ratio $b/a$, and effective
surface brightness $I_e$ (all as measured from SDSS imaging data for a
de~Vaucouleurs galaxy model).  Solid black lines show the cumulative
distribution in rank, solid gray lines show the null hypothesis for
this distribution, and vertical dash-dot lines show the location of
the maximum difference between the two distributions.}
\end{figure*}

\section{DISCUSSION AND IMPLICATIONS FOR FUTURE SURVEYS}\label{sec:future}

The large number of new strong lenses confirmed by the SLACS Survey
demonstrates the technical feasibility of carrying out spectroscopic
surveys for strong gravitational lenses, and suggests that similar
gravitational lens surveys should be considered as explicit goals of
future large-scale spectroscopic surveys.  The SLACS results are
particularly notable when one considers that this type of lens survey
was never one of the science or design goals of the SDSS, and the
survey is thus a testament to the performance of the SDSS telescope
and spectrograph, and to the optimal quality of the SDSS spectroscopic
data pipeline.  With the clarity of hindsight, we may identify those
particular features of the SDSS that were beneficial to the
spectroscopic lens survey project and those that could conceivably be
improved in any future survey to increase the yield of
spectroscopically selected strong lenses suitable for specified
science goals.

The large parent sample of galaxies observed is a clear advantage
afforded by the SDSS for this project or any other rare object survey.
Empirically, we have found that spectroscopic lenses and lens
candidates appear with a frequency on the order of one in a thousand
in the SDSS: starting with $\sim10^5$ early-type galaxies is therefore
one key to success.  At the same time, the large number of highly
magnified galaxies that we have found close to the spectroscopic
detection limit (note the generally low emission-line signal-to-noise
ratio in the discovery spectra of Figure~\ref{specfig}) suggests that
we may only be seeing the tip of the iceberg.  We conclude that (1) a
spectroscopic lens survey will always be at an advantage if it can
``piggy back'' on a more broadly conceived spectroscopic survey that
does not have lenses as its sole scientific goal, and (2) future
spectroscopic surveys that go significantly deeper than the SDSS
should, all other factors being equal, discover a significantly larger
fraction of gravitational lenses.

One might suspect that a significant factor for SLACS survey is the
large (3--arcsec diameter) aperture of the SDSS spectroscopic fibers.
Certainly a larger fiber will always collect more photons, but for
lenses with image splittings of $\sim 2\arcsec$ (i.e.\ $\sim 1\arcsec$
Einstein radii) such as we have presented here, a 3--arcsec fiber
could afford an advantage over smaller fibers disproportionate to its
larger aperture due to the concentration of lensed flux away from the
center of the fiber.  We can test the importance of this aperture-size
effect by simulating observations with varying fiber apertures.  We
assume an Einstein radius of $1\farcs 2$ (an approximate median value
from Paper III) and an exponential-disk background galaxy with a disk
scale length of $0\farcs2$.  We compute lensed images for a range of
source-galaxy impact parameters using a singular isothermal sphere
lens model.  We then smear the images with $1\farcs3$ seeing (the
median SDSS spectroscopic seeing for our targets) and integrate over
the 3--arcsec diameter SDSS fiber.  If we reduce the fiber aperture to
2--arcsec in diameter, we find that the flux collected from the lensed
galaxy is reduced to a fraction of $0.4$--$0.5$, depending on impact
parameter.  This is essentially the factor of $(1/1.5)^2 \simeq 0.44$
corresponding to the diminished aperture size.  Thus for lenses of the
angular size typical of SLACS lenses, the advantage of a large fiber
is just the simple advantage of a large aperture due to the smearing
effects of seeing; a smaller fiber would not have lost a
disproportionate amount of line flux due to lensing geometry.
However, for lenses with larger image splittings (which would occur
for higher redshift sources behind the same foreground galaxies), a
small fiber would likely be a qualitative disadvantage to a
spectroscopic lens survey. Because a fiber-diameter much greater than
3 arcsec would probably increase spurious non-lensed interlopers, we
conclude that the SDSS fiber-size represents a very good compromise,
when searching for massive early-type lens galaxies at these
redshifts.

For two principal reasons, our spectroscopic survey owes much success
to having targeted lens candidates with background redshifts confirmed
by multiple emission lines.  First, the incidence of false-positive
spectroscopic detections and emission-line mis-identifications is
negligible.  Second, to fixed line-flux limits, intermediate-redshift
emission from \oii, \oiiib, and the hydrogen Balmer series is more
common than Lyman-$\alpha$ emission at high redshift
\citep{cadis_hippelein, cadis_maier}, and thus we see an abundance of
oxygen/Balmer lenses.  However, for optical surveys, this survey
strategy limits the background redshifts to $z_{\mathrm{BG}} \lesssim
0.8$, beyond which \hb\ and \oiiib\ move out of the observable band.
There is a definite incentive to discover significant numbers of
lenses with higher lens and source redshifts, both to probe evolution
in the lens population and to observe lensed images at larger physical
radii within the lens galaxy in order to obtain greater leverage on
dark-matter halos.  Therefore in designing future surveys to discover
gravitationally lensed emission-line galaxies at higher redshift, two
obvious considerations are increased survey depth to detect a fainter
population, and increased spectroscopic resolution beyond the $\lambda
/ \Delta \lambda \approx 1800$ resolution of the SDSS in order to
split the \oii\ doublet and resolve the characteristic asymmetric
profile of Lyman-$\alpha$ emission (thus permitting more secure
single-line redshifts).

One of the greatest uncertainties at the outset of the SLACS survey
was simply the fraction of strong lenses within the candidate list.
One can attempt to calculate the lensing probabilities of the
candidates given $z_{\mathrm{FG}}$, $z_{\mathrm{BG}}$, $\sigma_a$,
assuming a model for the SDSS spectroscopic observations, and assuming
a luminosity function (LF) and size/shape for the background
emission-line galaxies, as was done in \citet{bolton_speclens}, but
the ingredients are all uncertain.  To put it simply, we did not know
beforehand whether or not the majority of our targets would owe their
high-redshift line emission to the PSF-smeared wings of bright
unlensed galaxies at large impact parameter rather than to
Einstein-ring images of faint galaxies closely aligned with the SDSS
target galaxy.  The high ratios of lenses to non-lenses and of
quads/rings to doubles that we see are suggestive of a large
magnification bias in our selection, with highly magnified faint
lensed galaxies detected with greater frequency than less magnified
lensed galaxies or unlensed projections.  Future work will quantify
the extent of any such magnification bias, and will derive statistical
constraints on the lens and source populations based on the incidence
of lensing within the survey.

\section{CONCLUSIONS AND FUTURE WORK}\label{sec:conclusions}

The {\sl HST} Cycle--13 SLACS Survey\footnote{{\sl HST}\,program
\#10174; PI: Koopmans.} has produced a catalog of 19 previously
unknown early-type strong gravitational lens galaxies.  The $\ge68$\%
fraction of genuine lenses within the the first 28 candidates observed
shows that the survey strategy---spectroscopic candidate selection
from within the SDSS, followed by {\sl HST}-ACS Snapshot observations
of the systems with the largest estimated lensing cross sections---is
an efficient means for the discovery of new gravitational lenses;
similar lens surveys should be considered as an explicit goal of
future spectroscopic surveys.  By targeting galaxies with the highest
estimated lensing cross sections, the Cycle--13 SLACS survey has
effectively selected high-mass lensing galaxies (strong lensing cross
section is simply proportional to lens mass within the Einstein
radius).  The natural selection of lenses in other surveys is also
weighted by lensing cross section, and thus strong lenses are
typically massive galaxies.  During {\sl HST} Cycle--14, we will extend
our survey to galaxies of lower mass in order to use strong lensing
and stellar dynamics to measure the mass dependence of early-type
galaxy structure and mass-to-light ratio\footnote{{\sl HST}\,program
\#10587; PI: Bolton; 118 Snapshot targets.}.  These galaxies will have
a lower lensing cross section and hence should have a lower lensing
rate, although magnification bias may skew the distribution of
candidates in favor of lenses over non-lenses.  The final combined
lens sample will be a unique resource for the detailed measurement of
the mass profile of early type galaxies within the effective radius.

The SLACS sample represents the largest single catalog of uniformly
selected early-type gravitational lens galaxies assembled to date.
The lensed images provide an aperture mass constraint within a typical
scale of $\sim R_E / 2$.  The SLACS lens galaxies are all brighter
than their lensed background galaxies by many magnitudes, and are thus
ideally suited to detailed photometric measurement as with the current
ACS data.  They are also excellent candidates for spatially resolved
spectroscopy to constrain dynamical galaxy models in combination with
the mass constraints from strong lensing.  In fact, nearly all of the
SLACS galaxies already have well-measured luminosity-weighted velocity
dispersions inside the 3--arcsec SDSS fiber (see Table~\ref{galtab}),
and thus we have already nearly tripled the number of known
gravitational lenses with known stellar velocity dispersions.  Paper
III will present the constraints on the logarithmic slope of the total
radial density profile (luminous plus dark matter) that can be
obtained by combining lens models with the aperture-integrated SDSS
velocity dispersions.  We may derive tighter constraints and a
significant decomposition into luminous and dark components by
obtaining spatially resolved line-of-sight velocity dispersion
measurements within the lens galaxies (see KT).

An analysis of the distribution of SLACS lens-galaxy photometric and
structural parameters within control samples from the SDSS database
shows that the SLACS lenses are typical of their SDSS parent sample
with regard to color and ellipticity. The location of SLACS lens
galaxies within the FP is the subject of Paper II. However, the SLACS
lens galaxies exhibit a somewhat significant bias towards brighter and
higher surface-brightness galaxies at fixed redshift and velocity
dispersion (\S\,\ref{sec:stats}; see also paper II). The significance
of this bias can be confirmed with a larger sample of similarly
selected lenses, which we plan to obtain with the continuation of our
ACS survey during {\sl HST} Cycle--14.

\acknowledgments

Results in this paper are based in part on data from the Sloan Digital
Sky Survey (SDSS) archive. Funding for the creation and distribution
of the SDSS Archive has been provided by the Alfred P. Sloan
Foundation, the Participating Institutions, the National Aeronautics
and Space Administration, the National Science Foundation, the
U.S. Department of Energy, the Japanese Monbukagakusho, and the Max
Planck Society. The SDSS Web site is {\tt http://www.sdss.org/}.  The
SDSS is managed by the Astrophysical Research Consortium (ARC) for the
Participating Institutions. The Participating Institutions are The
University of Chicago, Fermilab, the Institute for Advanced Study, the
Japan Participation Group, The Johns Hopkins University, Los Alamos
National Laboratory, the Max-Planck-Institute for Astronomy (MPIA),
the Max-Planck-Institute for Astrophysics (MPA), New Mexico State
University, University of Pittsburgh, Princeton University, the United
States Naval Observatory, and the University of
Washington.

These results are also based in part on
observations obtained with the 6.5-m Walter Baade
telescope of the Magellan Consortium at Las Campanas Observatory.

These results are also based in part on
observations obtained under program GN-2004A-Q-5 at the Gemini
Observatory, which is operated by the Association of Universities for
Research in Astronomy, Inc., under a cooperative agreement with the
NSF on behalf of the Gemini partnership: the National Science
Foundation (United States), the Particle Physics and Astronomy
Research Council (United Kingdom), the National Research Council
(Canada), CONICYT (Chile), the Australian Research Council
(Australia), CNPq (Brazil) and CONICET (Argentina).

These results are also based in part on observations made
with the NASA/ESA Hubble Space Telescope, obtained at the Space
Telescope Science Institute (STScI), which is operated by the
Association of Universities for Research in Astronomy, Inc., under
NASA contract NAS 5-26555.  These observations are associated with
program \#10174.  Support for program \#10174 was provided by NASA
through a grant from STScI.

The work of LAM was carried out at Jet
Propulsion Laboratory, California Institute of Technology, under a
contract with NASA.

The authors thank the anonymous referee for a careful reading
of the manuscript and for a constructive report.

\appendix
\section{B-SPLINE GALAXY MODEL SUBTRACTION}
\label{bsplineapp}

The b-spline technique is a well-known method for fitting a
piecewise-defined polynomial of arbitrary order to the dependence of a
series of data values upon an independent variable
\citep[e.g.][]{deboor_bsplines}.  The coefficients of the polynomial
change at breakpoints in the independent-variable domain, whose
spacing may be chosen to allow more or less freedom depending upon the
level of detail to be fit.  A b-spline of order $n$ (where by
convention $n=1$ is piecewise constant, $n=2$ is linear, $n=3$ is
quadratic, $n=4$ is cubic, and so on) has continuous derivatives to
order $n-2$ across the breakpoints.  For our galaxy modeling we use
b-splines of order $n=4$, which have continuous 0th, 1st, and 2nd
derivatives.  The coefficients of the b-spline are determined by
fitting to the data in a least-squares sense, given the breakpoint
spacing and derivative-continuity conditions.  A b-spline model may be
reinterpreted in terms of a number of localized basis functions within
the domain whose shapes are set by the order of the b-spline and by
the breakpoint spacing and whose amplitudes are determined by the fit
to the data.  The fit itself is entirely linear, and the localized
influence of each basis function within the domain implies that only a
banded-diagonal matrix need be inverted in the solution for the
coefficients.

The radial b-spline technique for galaxy images permits smooth fitting
of arbitrary radial brightness profiles.  Since we do not have
multiple dithered exposures to combine, we perform our b-spline
galaxy-model fitting in the native ACS pixel coordinates of the
images.  We account for the distortion in the ACS by using the
solution provided in the image headers to compute relative
tangent-plane RA and Dec values for all pixels in the image, which are
taken as the independent variables for the fit.  We fit for surface
brightness as a function of position using cosmic-ray-masked
flat-fielded images (i.e.\ calibrated by the CALACS software
pipeline\footnote{ (see {\tt http://www.stsci.edu/hst/acs/})} to
measure surface brightness rather than counts-per-pixel), so our fits
are not biased by slight variations in pixel area across the images.
The one non-linear step in our b-spline galaxy-model fitting is the
determination of the center of the lensing galaxy; we describe our
centering method further below.  The model-fitting procedure is
carried out using the adopted center for the lens galaxy within a
suitable subsection of the ACS field (typically $12\arcsec \times
12\arcsec$), along with an error image, a mask specifying cosmic-ray
and other zero-weight pixels, and a mask corresponding to stars,
neighboring galaxies, and any apparent background-galaxy features.
The cosmic-ray masks are initially generated by the LACOSMIC software
\citep{van_dokkum_lacosmic} and adjusted manually over the image
subsection surrounding the lens; the neighboring-object masks are
created manually.  For each pixel in the sub-image, the radial offset
$R$ from the galaxy center (in arcseconds) and the azimuthal angle
$\theta$ relative to a fixed position angle are computed.  A set of
breakpoints in $R$ is chosen for the fit, typically every $0\farcs
2$--$0\farcs 3$ in the central $1\arcsec$ to $2\arcsec$, with
increased spacing further out.  (We note again that the
ACS pixel scale is 0\farcs05 per pixel.)
For the case of complete Einstein
rings, breakpoints at the ring radius can be removed to allow smoother
interpolation of the lens-galaxy model over the ring region.  A
multipole angular dependence is incorporated into the fit as follows.
If the purely one-dimensional b-spline fit is represented as
\begin{equation}
I(R) = \sum_k a_k f_k (R) ~~,
\end{equation}
where the $f_k(R)$ are the localized basis functions that
are only non-zero over a small range of breakpoints and
the $a_k$ are their amplitudes, then the two-dimensional
fit is represented by
\begin{equation}
I(R, \theta) = \sum_{m,k} \left[b_{mk} \cos(m \theta)
+ c_{mk} \sin(m \theta) \right] f_k (R) ~~.
\end{equation}
The number of multipole orders to be fit is chosen individually for
each galaxy, with $m=0$ (monopole) and $m=2$ (quadrupole) always
present and higher-order terms added if necessitated by systematic
angular structure in the residual images.  With the occasional
exception of an $m=1$ (dipole) term to capture slightly disturbed
morphologies or mild mis-centering in the central regions, only even
multipole orders are used.  This form of angular dependence is ideal
for fitting early-type galaxies: the fit remains linear, the global
symmetries seen in early-type galaxies are naturally reflected in the
low-order terms, and effects such as isophotal twists, varying
ellipticity with radius, and diskiness/boxyness can be captured with
minimal effort (with only the last of these effects requiring
multipole orders beyond the quadrupole).  We note also that the
monopole term can automatically include a fit to the sky background.

Following the completion of the initial b-spline model-fitting step
(which for our $12\arcsec \times 12\arcsec$, 240$\times$240-pixel
images generally takes 2--3 seconds on a 2.53GHz Pentium 4 Linux PC),
we examine the residual (data$-$model) image for faint features not
associated with the smooth galaxy model and not initially masked, and
perform a second fit with an updated mask, also adding multipole
orders as needed.  Since at this stage we are principally concerned
with generating high-quality residual images and not with the
measurement of lens-galaxy parameters, we fit directly to the images
without convolving the b-spline model with the {\sl HST} point-spread
function.  As an example, we describe the fit to the sky-subtracted
image of the E3 lens galaxy SDSSJ0912$+$0029 to give the F814 residual
image shown in Figure~\ref{acsfig}, which uses breakpoints spaced
every $0\farcs 3$ in $R$ and multipole terms of order 0, 2, and 4
(monopole, quadrupole, and octopole).  The ratio of the $m=2$
amplitudes (defined as $\sqrt{b_{2k}^2 + c_{2k}^2}$) to the $m=0$
amplitudes rises from $\approx 0.25$ in the center to $\approx 0.4$ at
the effective radius (about $3\arcsec$), and the ratio of the $m=4$ to
the $m=0$ amplitudes rises from $\approx 0$ to $\approx 0.1$ over the
same range.  Some of this increase in angular structure with radius is
due to our fitting directly to the PSF-blurred data.

The centering necessary for the b-spline galaxy fit is accomplished by
first fitting with respect to a best-guess center with monopole,
dipole, and quadrupole terms.  This fit is then evaluated in a ring
and used to determine a flux centroid, which is adopted as a new
best-guess center for another iteration of the same procedure.  Within
a few iterations this process---which amounts to minimizing the dipole
term---converges to a position that we adopt as the constrained center
for the final model fit.

\section{NOTES ON INDIVIDUAL SYSTEMS}
\label{maybeapp}

\subsection{Lenses}

\noindent \underline{SDSSJ0216$-$0813:} A faint but definite counterimage
is seen to the NE in both bands in this system, opposite the
more obvious extended arc.

\medskip

\noindent \underline{SDSSJ0912$+$0029:} This system shows a faint extended
counter-arc to the S in the F814W band, opposite the more
prominent Northern arc.  Both images are only marginally
detected in the $F435W$ band.

\medskip

\noindent \underline{SDSSJ0956$+$5100:} This system includes a small round red
companion to the N of the lens galaxy, not to be confused
with the prominent lensed images.

\medskip

\noindent \underline{SDSSJ1251$-$0208:} This galaxy has extended spiral
structure in addition to the lensing bulge.

\medskip

\noindent \underline{SDSSJ1330$-$0148:} This system is judged to be a lens
on the basis of the compact counterimage detected at high
significance in both bands.

\medskip

\noindent \underline{SDSSJ1402$+$6321:} This lens, with faint quadruple
images, was the first system observed by the SLACS survey and is
the subject of \citet{bolton_1402}.

\medskip

\noindent \underline{SDSSJ1618$+$4353:} The angular resolution of {\sl HST}
reveals this system to be a pair of foreground galaxies.  The pair
lenses a compact background source into a 3$+$1 quad configuration.
The extra compact image to the west of the counterimage is more blue
in color than the lensed images and thus does not pertain to the
lensed configuration.

\medskip

\noindent \underline{SDSSJ1718$+$6424:} A second nearby galaxy contributes
significantly to the lensing potential in this system.

\subsection{Other Systems}

\noindent \underline{SDSSJ1117$+$0534, SDSSJ1259$+$6134:}
The imaging depth is not great enough to 
unambiguously show whether the putative lensed
features appear in both bands, but 
the general geometry is plausibly similar to J0912+0029, which has an 
extremely solid gravitational lens model (Paper III). 

\medskip

\noindent \underline{SDSSJ1636$+$4707:}
The curvature of the arcs is suggestive of ring geometry, 
and the knot-counterknot features along the North-South axis may be 
images of each other.  Deeper observations would reveal whether or not
this system is a real Einstein ring,
but it is not certain from the current data.

\medskip

\noindent \underline{SDSSJ1702$+$3320:}
Both F435W and F814W data show possible 
counter-image features with plausible relative orientations and 
separations at low signal-to-noise.

\end{document}